\documentclass[aps,prb,superscriptaddress,reprint,showpacs,twocolumn,preprintnumbers,floatfix]{revtex4-2}
\usepackage[dvips]{graphicx}
\usepackage[latin1]{inputenc}
\usepackage{xcolor,bm}
\usepackage{multirow,easyReview}
\usepackage[charter]{mathdesign}
\usepackage{hyperref}
\hypersetup{
	colorlinks=true,
	linkcolor=blue,
	citecolor=blue,
	urlcolor=blue
}
\begin{document}
	\draft
	
	\hyphenation{a-long}
	
	\title{Complex vortex-antivortex dynamics in the magnetic superconductor EuFe$_{2}$(As$_{0.7}$P$_{0.3}$)$_{2}$}
	
	\author{Giacomo Prando}
	\thanks{These two authors contributed equally}
	\affiliation{Dipartimento di Fisica, Universit\`a degli Studi di Pavia, Pavia 27100, Italy}
	\author{Daniele Torsello}
	\thanks{These two authors contributed equally}
	\affiliation{Department of Applied Science and Technology, Politecnico di Torino, Torino 10129, Italy}
	\affiliation{Istituto Nazionale di Fisica Nucleare, Sezione di Torino, Torino 10125, Italy}
	\author{Samuele Sanna}
	\affiliation{Dipartimento di Fisica e Astronomia ``A. Righi'', Universit\`a di Bologna, Bologna 40127, Italy}
	\author{Michael J. Graf}
	\affiliation{Physics Department, Boston College, Chestnut Hill, Massachusetts 02467, USA}
	\author{Sunseng Pyon}
	\affiliation{Department of Applied Physics, The University of Tokyo, Hongo, Bunkyo-ku, Tokyo 113-8656, Japan}
	\author{Tsuyoshi Tamegai}
	\affiliation{Department of Applied Physics, The University of Tokyo, Hongo, Bunkyo-ku, Tokyo 113-8656, Japan}
	\author{Pietro Carretta}
	\affiliation{Dipartimento di Fisica, Universit\`a degli Studi di Pavia, Pavia 27100, Italy}
	\author{Gianluca Ghigo}
	\thanks{Corresponding author: gianluca.ghigo@polito.it}
	\affiliation{Department of Applied Science and Technology, Politecnico di Torino, Torino 10129, Italy}
	\affiliation{Istituto Nazionale di Fisica Nucleare, Sezione di Torino, Torino 10125, Italy}
	
	%%%%%%%%%%%%%%%%%%%%
	\widetext
	%%%%%%%%%%%%%%%%%%%%
	
	%%%%%%%%%%%%%%%%%%%%%%%%%%%%%%%%%%%%%%%%%%%%%%%%%%%%%%%%%%%%%%%%%%%%%%%%%%%%%%%%
	\begin{abstract}
		We report on the investigation of the magnetic superconductor EuFe$_{2}$(As$_{0.7}$P$_{0.3}$)$_{2}$ based on muon-spin spectroscopy and ac magnetic susceptibility ($\chi$) measurements. The dependence of the internal field at the muon site on temperature is indicative of a ferromagnetic ordering of Eu$^{2+}$ magnetic moments and only the conventional magnon scattering governs the longitudinal relaxation rate at low temperatures. At the same time, we observe a rich phenomenology for the imaginary component of the susceptibility $\chi^{\prime\prime}$ by means of both standard ac susceptibility and a novel technique based on a microwave coplanar waveguide resonator. In particular, we detect activated trends for several features in $\chi^{\prime\prime}$ over frequencies spanning ten orders of magnitude. We interpret our results in terms of the complex dynamics of vortices and antivortices influenced by the underlying structure of magnetic domains.
	\end{abstract}
	%%%%%%%%%%%%%%%%%%%%%%%%%%%%%%%%%%%%%%%%%%%%%%%%%%%%%%%%%%%%%%%%%%%%%%%%%%%%%%%%
	
	\date{\today}
	
	\maketitle
	
	%%%%%%%%%%%%%%%%%%%
	\narrowtext
	%%%%%%%%%%%%%%%%%%%%
	
	\section{Introduction}\label{SectIntro}
	In Eu-based pnictide superconductors, the superconducting and magnetic orders coexist over a large temperature range, with magnetism arising from the large Eu$^{2+}$ local moments in the spacing layers between the Fe-As planes. A wide variety of behaviours are observed in these materials depending on the specific structure, doping level \cite{Ren2009PRB,Cao2011JPCM,Zapf2013PRL}, applied pressure \cite{Miclea2009PRB,Tokiwa2012PRB}, and amount of quenched disorder \cite{Ghigo2020sust,Ghimire2021Materials} and range from re-entrant superconductivity \cite{Miclea2009PRB} to helical magnetic order \cite{Liu2016PRB,Smylie2019PRB,Koshelev2019PRB} and spontaneous formation of vortex-antivortex pairs \cite{Stolyarov2018sciadv,Vinnikov2019JETP,Devizorova2019PRL}. Optimally-doped EuFe$_2$(As$_{1-x}$P$_{x}$)$_2$ is particularly interesting because, upon decreasing temperature, it shows a transition to the superconducting state at the critical temperature $T_{SC}$ soon followed by the ordering of Eu$^{2+}$ magnetic moments to a canted antiferromagnetic state with a net ferromagnetic (FM) component along the crystallographic $c$-axis \cite{Xiao2009PRB,Zapf2011PRB,Nandi2014prb2}. Below the onset of the ordered magnetic state at the critical temperature $T_{M}$, a so-called striped Meissner domain structure is formed where parallel regions of Meissner domains alternate the direction of circulating currents. At lower temperatures, vortex-antivortex pairs nucleate spontaneously and populate domains with opposite magnetization, changing their shapes and widths \cite{Nandi2014PRB,Stolyarov2018sciadv}. Overall, Eu-based pnictides are an ideal playground to study the interplay between magnetism and superconductivity \cite{Zapf2016RPP}. 
	
	As a result of its complex underlying physical properties, EuFe$_2$(As$_{1-x}$P$_{x}$)$_2$ shows a rich phenomenology of dynamical processes whose nature is not fully clear at the moment. In particular, the dynamics may arise from Eu$^{2+}$ magnetic moments, magnetic domains, superconducting vortices, or a complex combination of these intertwined elements \cite{Ghigo2019PRR}. In this work, we clarify these issues by characterizing the magnetic and superconducting state of optimally-doped EuFe$_2$(As$_{1-x}$P$_{x}$)$_2$ by means of muon-spin spectroscopy ($\mu$SR) and complex ac susceptibility ($\chi$) measurements in a frequency range $\nu \simeq 0.5$ Hz $- 8$ GHz using a combination of conventional ac susceptometry and a novel technique based on a microwave (MW) coplanar waveguide resonator. $\mu$SR measurements are strongly influenced by the magnetic state of Eu$^{2+}$ moments and evidence a conventional magnon-driven relaxation mechanism with no additional contributions well below $T_{M}$. At the same time, we find evidences of two dissipation peaks in the imaginary component of the magnetic susceptibility $\chi^{\prime\prime}$ characterized by activated behaviours over a ten-orders-of-magnitude frequency range. We propose a plausible interpretation of our results in terms of the dynamics of spontaneously-generated vortices and antivortices within the ferromagnetic background arising from Eu$^{2+}$ magnetic moments, highlighting the different contributions from depinning processes within magnetic domains and those involving the tunneling of fluxoids between different domains.
	
	\section{Experimental details}\label{SectExp}
	
	\subsection{Samples preparation}
	EuFe$_2$(As$_{1-x}$P$_x$)$_2$ ($x = 0.3$) polycrystalline pellets were prepared by using EuAs, Fe$_2$As, and Fe$_2$P as precursors. Stoichiometric amounts of precursors were placed in an alumina crucible, which was sealed in a stainless steel container. The whole assembly was placed in a box furnace and heated at $1000^\circ$C for $20$-$30$ h with or without intermediated grindings. Finally, the reacted materials were ground into fine powder, made into pellets, put in an alumina crucible, sealed in stainless steel container, and fired at $1000^\circ$C for $10$ h. %We used Eu ingot (3N), Fe powder (3N), As pieces (7N), and P grains (6N) as starting materials for precursors. EuAs was prepared by directly reacting Eu and As in an evacuated quartz tube at 500$^\circ$C for 5-10 h followed by firing at 700$^\circ$C for 20-52 h with or without an intermediate grinding. For Fe$_2$P, Fe powder and P grains were sealed in an evacuated quartz tube and heated slowly up to 600$^\circ$C spending more than 24 h, followed by the reaction at 600$^\circ$C for 10 h. Fe$_2$As was prepared by sealing Fe powder and As pieces in an evacuated quartz tube, and heated at 500$^\circ$C for 10 h followed by the reaction at 700$^\circ$C for 24 h. 
	All handlings involving Eu-containing materials were done in a high-purity Ar-filled glove box to avoid oxidation of the material. The purity of the obtained samples was checked by X-ray diffraction.
	
	\subsection{Muon-spin spectroscopy}
	We used $\mu$SR to investigate the magnetic response of the EuFe$_{2}$(As$_{0.7}$P$_{0.3}$)$_{2}$ sample from a microscopic viewpoint. We performed the measurements on the General Purpose Surface-muon (GPS) spectrometer of the Swiss Muon Source at the Paul Scherrer Institute (Switzerland) and on the high-field (HiFi) spectrometer of the ISIS Neutron and Muon Source at the Rutherford Appleton Laboratories (UK) \cite{Muons}. In a $\mu$SR experiment, millions of spin-polarized positive muons are implanted in the investigated material and the evolution of their spin polarization in time is monitored under the effect of the local internal magnetic fields \cite{Blundell,YaouancDeReotier}. This is made possible by a space- and time-resolved detection of the positrons resulting from the decay of the muons (average lifetime $\langle\tau_{\mu}\rangle \sim 2.2 \; \mu$s) and emitted preferentially along the direction of the muon spin at the decay moment. In particular, the so-called asymmetry ($A$) quantifies the difference in the positron countings in two detectors oppositely located with respect to the sample and is directly proportional to the time autocorrelation function of the spin polarization. We performed the measurements on GPS without any applied magnetic field (zero field, ZF) in conditions of active compensation of the Earth's magnetic field. We also probed the effect of external magnetic fields applied parallel to the muon spin polarization at the moment of implantation (longitudinal field, LF) with $H_{LF} \leq 35$ kOe on HiFi.
	
	\subsection{ac susceptibility}
	We measured the magnetic ac susceptibility $\chi$ of EuFe$_{2}$(As$_{0.7}$P$_{0.3}$)$_{2}$ loose powders (obtained after grinding the pellets) with frequency up to 1 kHz using a Quantum Design MPMS3 commercial susceptometer. We focused on the dependence of $\chi$ on temperature upon both warming and cooling protocols. In the following discussion, we refer only to data obtained in the absence of a polarizing magnetic field ($H_{dc} = 0$ Oe) and with a fixed value $H_{ac} = 1$ Oe for the alternating magnetic field.
	
	The high-frequency ($\nu \simeq 7.9$ GHz) susceptibility of EuFe$_{2}$(As$_{0.7}$P$_{0.3}$)$_{2}$ powder fragments was investigated by means of a MW coplanar waveguide resonator (CPWR) technique \cite{Torsello2019PRB,Torsello2020PRAppl}. The sample under study was coupled to a CPWR, in a region where the MW magnetic field is quite homogeneous and its amplitude is $H_{ac} \simeq 1$ Oe \cite{Ghigo2017PRB}. Within a resonator-perturbation approach, the fractional shifts of the resonance frequency $\nu_0$ and of the quality factor $Q$ of the resonator, due to the presence of the small sample, are related to the complex magnetic susceptibility of the sample itself through 
	\begin{equation}\label{eq.CPWR}
		2\frac{\Delta \nu_0}{\nu_0} + i \Delta\left(\frac{1}{Q}\right) \simeq -\left(\Gamma_f \chi' + i \Gamma_Q \chi''\right),
	\end{equation}
	where $\Gamma_f$ and $\Gamma_Q$ subsume the geometrical factors connected to the distribution of the MW fields around the resonator and to the sample geometry \cite{libro}. Two different geometrical factors need to be used for the real and imaginary parts of Eq.~\ref{eq.CPWR}, because the approximations consequence of the small-perturbation hypothesis affect differently the two terms. These factors could be determined through a calibration procedure from data above $T_{SC}$, where the sample is assumed to behave as a normal metal. In this case, for a sample with well defined geometry (e.g., thin platelet with the magnetic field parallel to its broad face) the geometrical factors can be obtained by data fitting \cite{Ghigo2017PRB,libro}. This yields the absolute values of the complex $\chi$ in the whole temperature range. When the sample under test does not meet adequate geometrical requirements, as for the powder fragments analyzed below, the exact geometrical factors remain unknown and $\chi$ can be represented only in arbitrary units. 
	
	\section{Results}\label{SectRes}
	
	\subsection{Muon-spin spectroscopy}
\begin{figure}[t!] 
		\centering
		\includegraphics[width=0.48\textwidth]{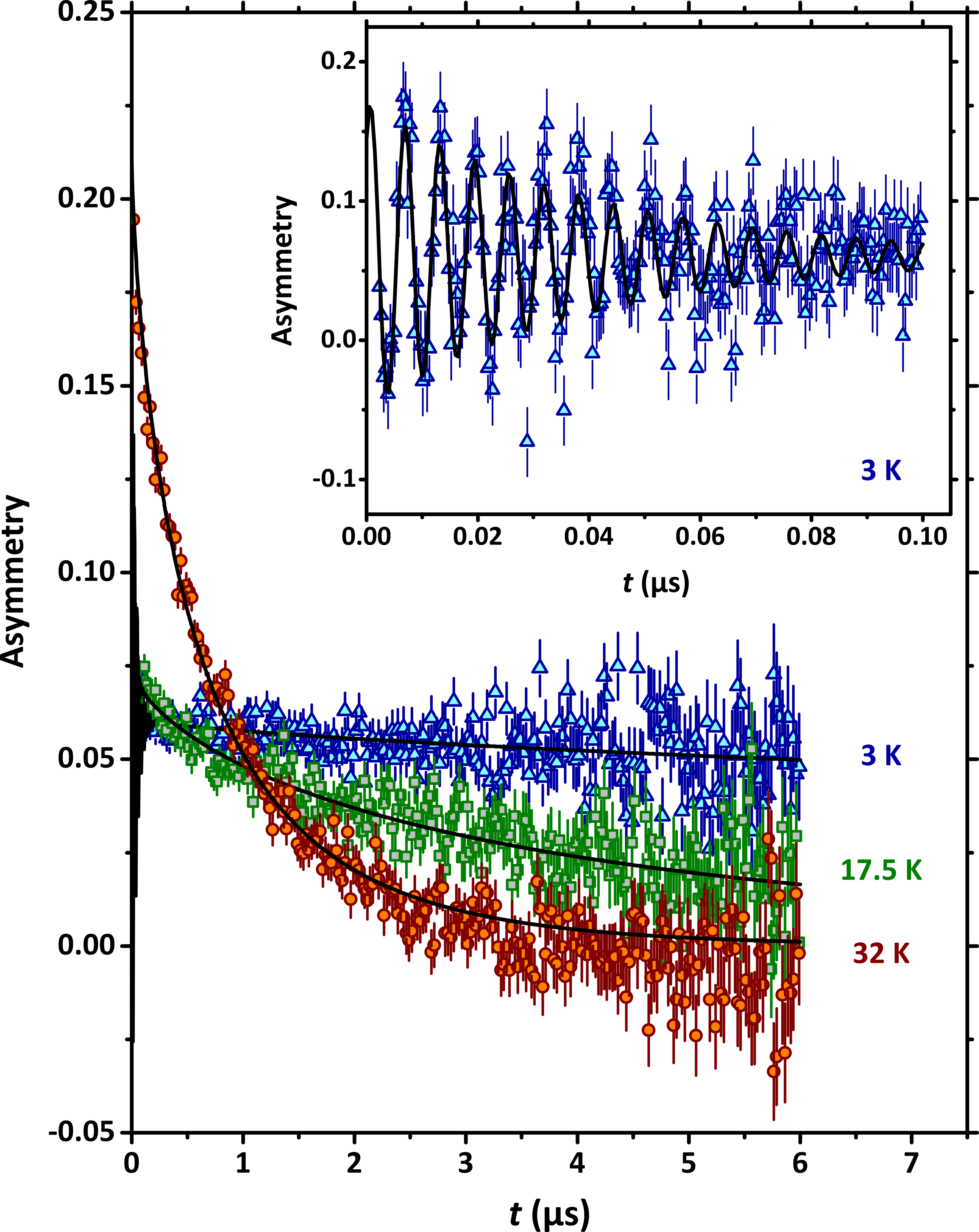}
		\caption{\label{FigRawDataMuSR} Main panel: dependence of the positron asymmetry as a function of the time elapsed after the muons implantation at representative temperatures in ZF conditions. The black curves are best-fitting functions according to Eq.~\ref{EqZFFittingFunction}. Inset: blow-up of the short-time limit highlighting the development of coherent oscillations at low temperatures. The oscillations are absent in the main panel because of the different choice for the binning factor.}
	\end{figure}

	The time evolution of the asymmetry in ZF conditions, reported in Fig.~\ref{FigRawDataMuSR} for selected temperatures, demonstrates the development of a phase transition to a long-range ordered magnetic phase over the whole sample volume. The high-temperature curve shows that the spin polarization is exponentially damped over a characteristic time of $\sim 1 \; \mu$s. On the other hand, fast coherent oscillations develop at low temperatures during the first $\sim 0.1 \; \mu$s of acquisition and, at the same time, the long-time behaviour is less damped if compared to what is observed at higher temperatures.
	
	\begin{figure}[t!] 
		\centering
		\includegraphics[width=0.48\textwidth]{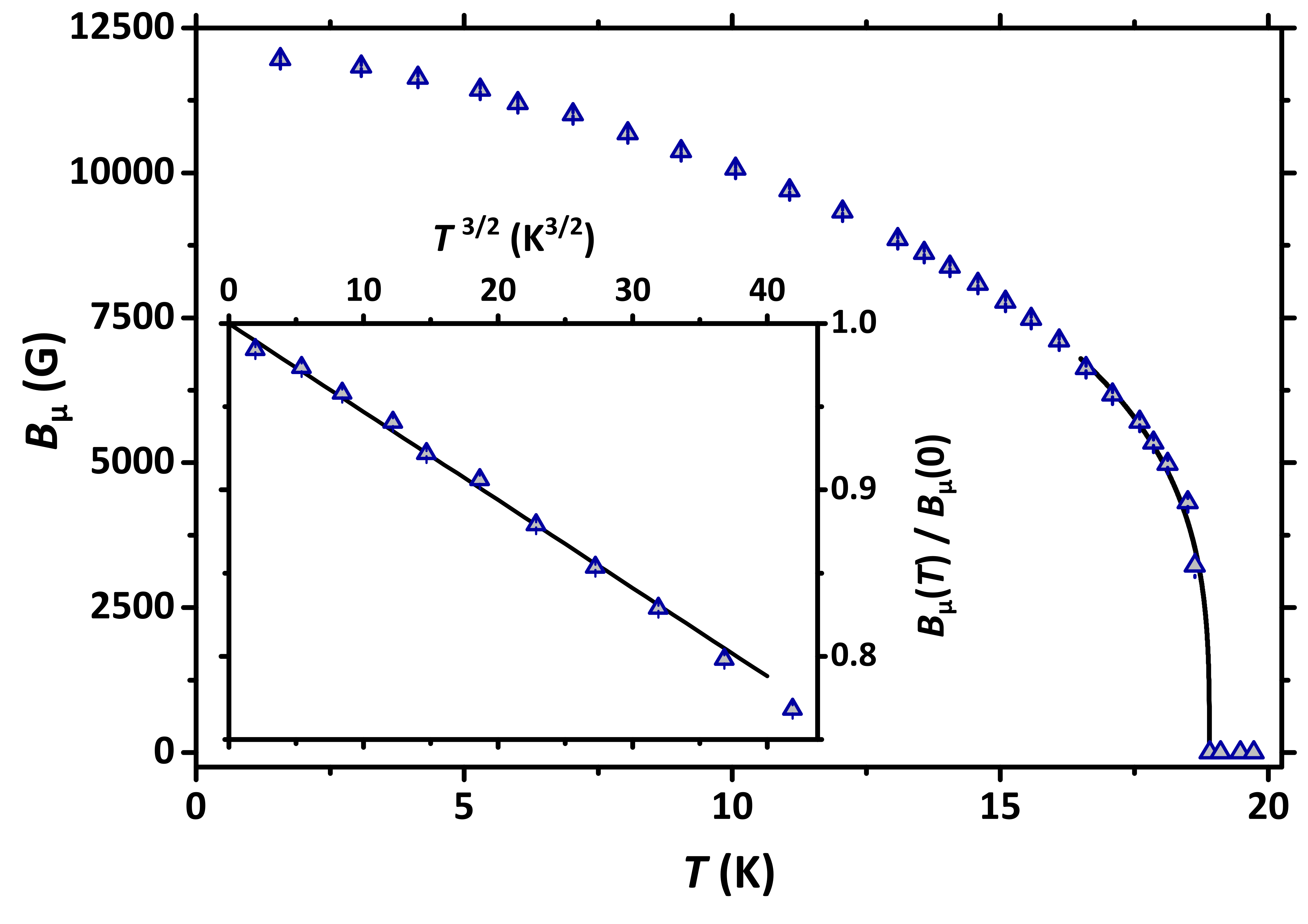}
		\caption{\label{FigMuSRfitNew1} Main panel: dependence of the average local magnetic field at the muon site on temperature. The continuous line is a best-fitting curve based on Eq.~\ref{EqOrderParameter}. Inset: blow-up of the average local magnetic field at the muon site at low temperatures normalized over its extrapolated value at zero temperature. The continuous line is a guide to the eye highlighting a dependence on $T^{3/2}$.}
	\end{figure}
	In order to describe these results in a more quantitative way, we refer to the following fitting function for the dependence of the asymmetry on time ($t$)
	\begin{equation}\label{EqZFFittingFunction}
		\frac{A(t)}{A_{0}} = a_{T} \cos\left(\gamma_{\mu} B_{\mu} t + \phi\right) e^{- \lambda_{T} t} + a_{L} e^{-(\lambda_{L} t)^{S}}
	\end{equation}
	over the entire investigated temperature window \cite{Eu227}. Here, $A_{0} \simeq 0.21$ is a spectrometer-dependent parameter defining the maximum initial value taken by the asymmetry. In the low-temperature magnetic state, $a_{T}$ and $a_{L}$ are the fractions of the signal coming from those muons probing static magnetic fields in transverse and parallel directions with respect to the spin polarization at the implantation moment, respectively. The average internal magnetic field at the muon site $B_{\mu}$ causes coherent oscillations in the precessing fraction $a_{T}$ whose angular frequency is determined via the muon gyromagnetic ratio $\gamma_{\mu} = 2\pi \times 13.554$ rad ms$^{-1}$ G$^{-1}$, while $\phi$ is a phase offset. Wider distributions of the local field value result in a higher damping of the coherent oscillations and in a higher value for the rate $\lambda_{T}$, in turn. On the other hand, the non-precessing fraction $a_{L}$ is sensitive to the fluctuations of the local magnetic field both above and below the critical temperature. These are quantified by the rate $\lambda_{L}$ which, in the fast-fluctuations limit, is directly proportional to the spin-lattice relaxation rate \cite{YaouancDeReotier}. Finally, the stretching exponent $S$ allows for a possible distribution of the longitudinal relaxation rate. The fitting function discussed above is specific to the GPS data taken in ZF conditions. It can be generalized straightforwardly to the HiFi data since, in that case, the worse time resolution associated with the pulsed nature of the muon beam makes it impossible to resolve the fast oscillations in the component, so that $a_{T}$ is kept fixed to $0$.
	
\begin{figure}[b!] 
		\centering
		\includegraphics[width=0.48\textwidth]{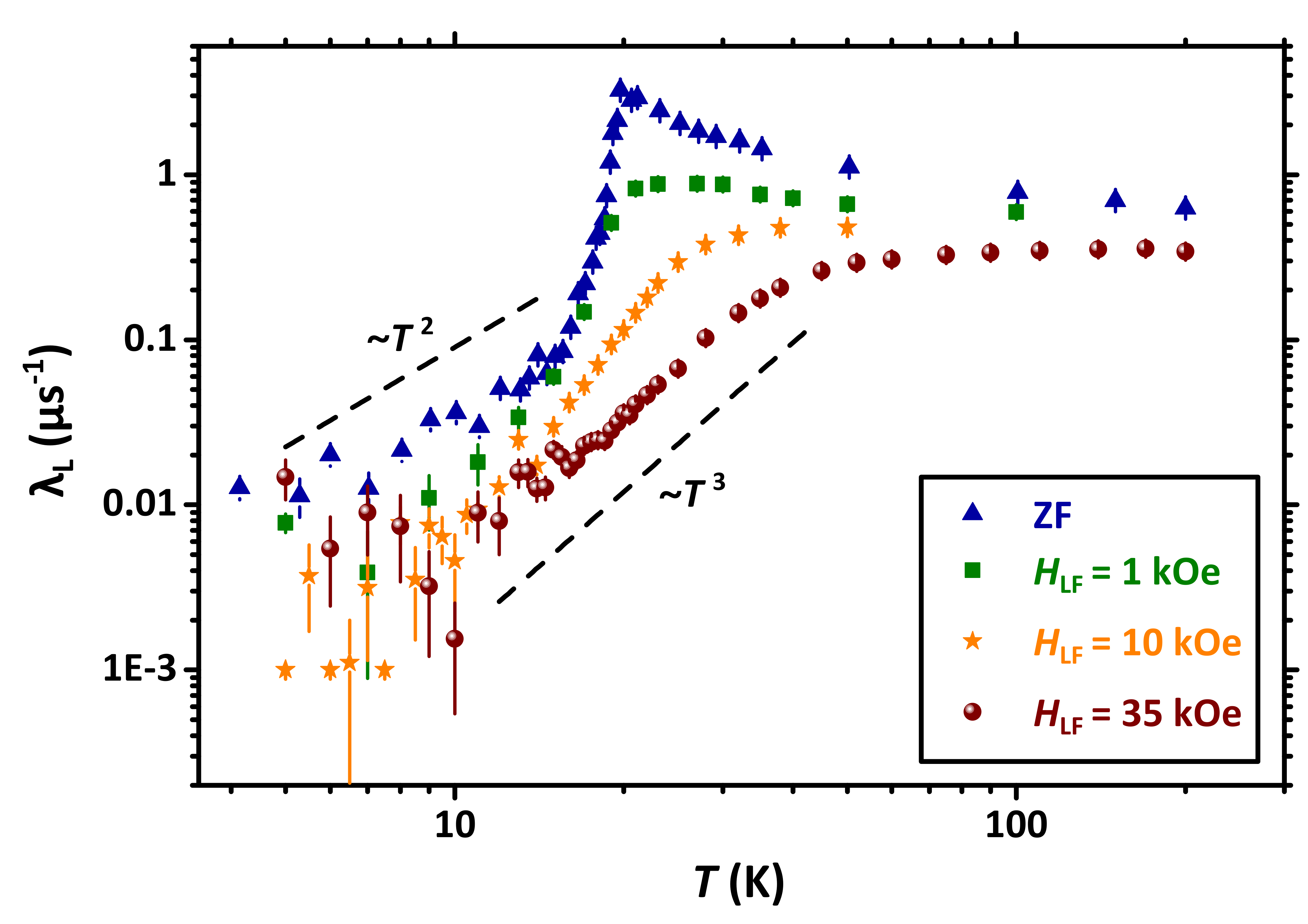}
		\caption{\label{FigMuSRfitNew2} Dependence of the longitudinal relaxation rate on temperature for different values of the applied LF. The dashed lines highlight the characteristic power-law dependence expected for a magnon-driven magnetic relaxation (see Sect.~\ref{SectDisc} later on).}
	\end{figure}

	The dependence of the main fitting parameters on temperature is reported in Figs.~\ref{FigMuSRfitNew1} and \ref{FigMuSRfitNew2}. The internal magnetic field at the muon site $B_{\mu}$ (Fig.~\ref{FigMuSRfitNew1}) shows the qualitative trend expected for an order parameter and, in particular, the following expression
	\begin{equation}\label{EqOrderParameter}
		B_{\mu}(T) = B_{\mu}(0) \left(1-\frac{T}{T_{M}}\right)^{\beta}
	\end{equation}
	can be used as a fitting function for $T \lesssim T_{M}$ as shown in the main panel of Fig.~\ref{FigMuSRfitNew1}. Based on the fitting results, we estimate $T_{M} = 18.90 \pm 0.05$ K for the critical temperature and $\beta = 0.3 \pm 0.01$ for the critical exponent characteristic of the order parameter. The inset to Fig.~\ref{FigMuSRfitNew1} shows a blow-up of the $B_{\mu}$ vs $T^{3/2}$ data in the opposite temperature limit $T \ll T_{M}$, highlighting a linear trend.
	
	The dependence of the longitudinal relaxation $\lambda_{L}$ on temperature under different conditions for the external magnetic field is reported in Fig.~\ref{FigMuSRfitNew2}. Above $T_{M}$, $\lambda_{L}$ approaches a high value of $\sim 1 \; \mu$s$^{-1}$ independent of the applied longitudinal magnetic field. We interpret this result as the effect of the fluctuating Eu$^{2+}$ magnetic moments in the paramagnetic regime. A well-defined critical peak is observed for $T \sim T_{M}$ in the ZF data which is progressively smeared upon increasing $H_{LF}$. Eventually, after a remarkably fast suppression of $\lambda_{L}$ for $T \lesssim T_{M}$, the data are consistent with a power-law dependence on temperature with characteristic exponent between $2$ and $3$ in the low-temperature regime $T \ll T_{M}$.
	
	\subsection{Magnetic ac susceptibility}
	
	\begin{figure}[t!] 
		\centering
		\includegraphics[width=0.48\textwidth]{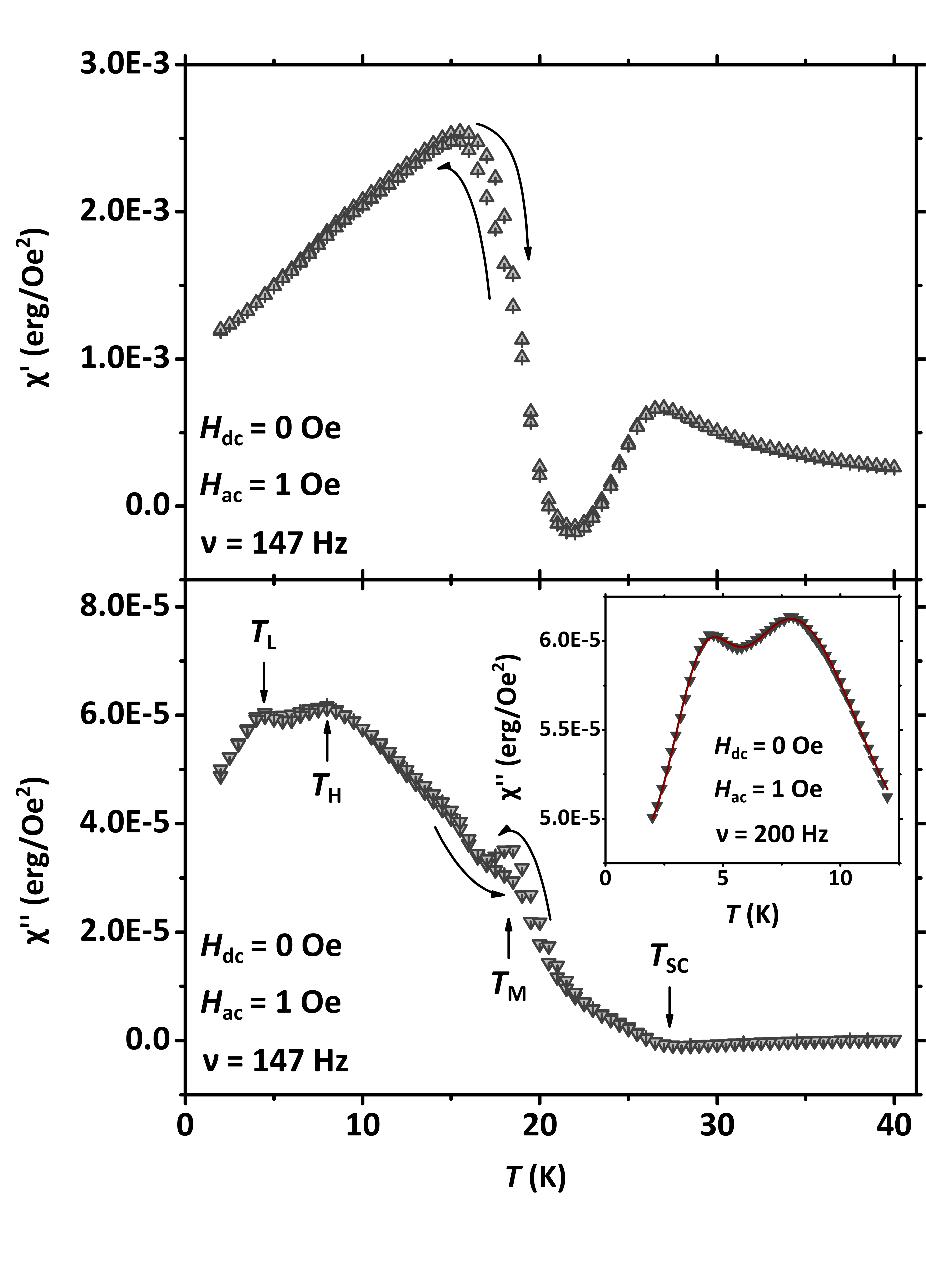}
		\caption{\label{FigReprAcSusc} Representative data showing the dependence of the real and imaginary components of the magnetic ac susceptibility of the sample on temperature (upper and lower panel, respectively). The curved arrows highlight a thermal hysteretic behaviour for the warming and cooling curves just below $19$ K. The straight arrows in the lower panel evidence the development of clear anomalies which we track more closely in dedicated measurements, as shown by the representative data in the inset to the lower panel. The continuous line in the inset is a fitting curve based on the empirical two-Gaussian expression discussed in Sect.~\ref{SectRes}.}
	\end{figure}
	Representative results for the real and imaginary components of $\chi$ at frequencies up to 1 kHz are reported in Fig.~\ref{FigReprAcSusc}. The development of the superconducting state extended over a bulk volume fraction leads to a sharp downturn at around $T_{SC} \simeq 27$ K in the real component of $\chi$ upon decreasing temperature superimposed to a Curie-like trend associated with a paramagnetic contribution from Eu$^{2+}$ magnetic moments (see the upper panel of Fig.~\ref{FigReprAcSusc}). The resulting maximum displays no dependence on either frequency or amplitude of the alternating field $H_{ac}$. Clear signs of thermal hysteresis can be observed just below the critical temperature to the magnetic phase $T_{M} \sim 19$ K, suggesting a first-order character of the phase transition. The shallow maximum developing in $\chi^{\prime\prime}$ at $T_{M} \sim 19$ K (see the lower panel of Fig.~\ref{FigReprAcSusc}) shows a negligible dependence on the frequency of $H_{ac}$. At lower temperatures, a maximum develops in $\chi^{\prime\prime}$ at $T_H \sim 8$ K and, eventually, an additional peak appears at around $T_L \sim 4$ K (see the inset to the lower panel of Fig.~\ref{FigReprAcSusc}). A phenomenological best-fitting procedure based on the sum of two Gaussian functions gives good results for all the experimental curves taken at $H_{ac} = 1$ Oe (see the inset to Fig.~\ref{FigReprAcSusc}). 
	
	\begin{figure}[t!] 
		\centering
		\includegraphics[width=0.48\textwidth]{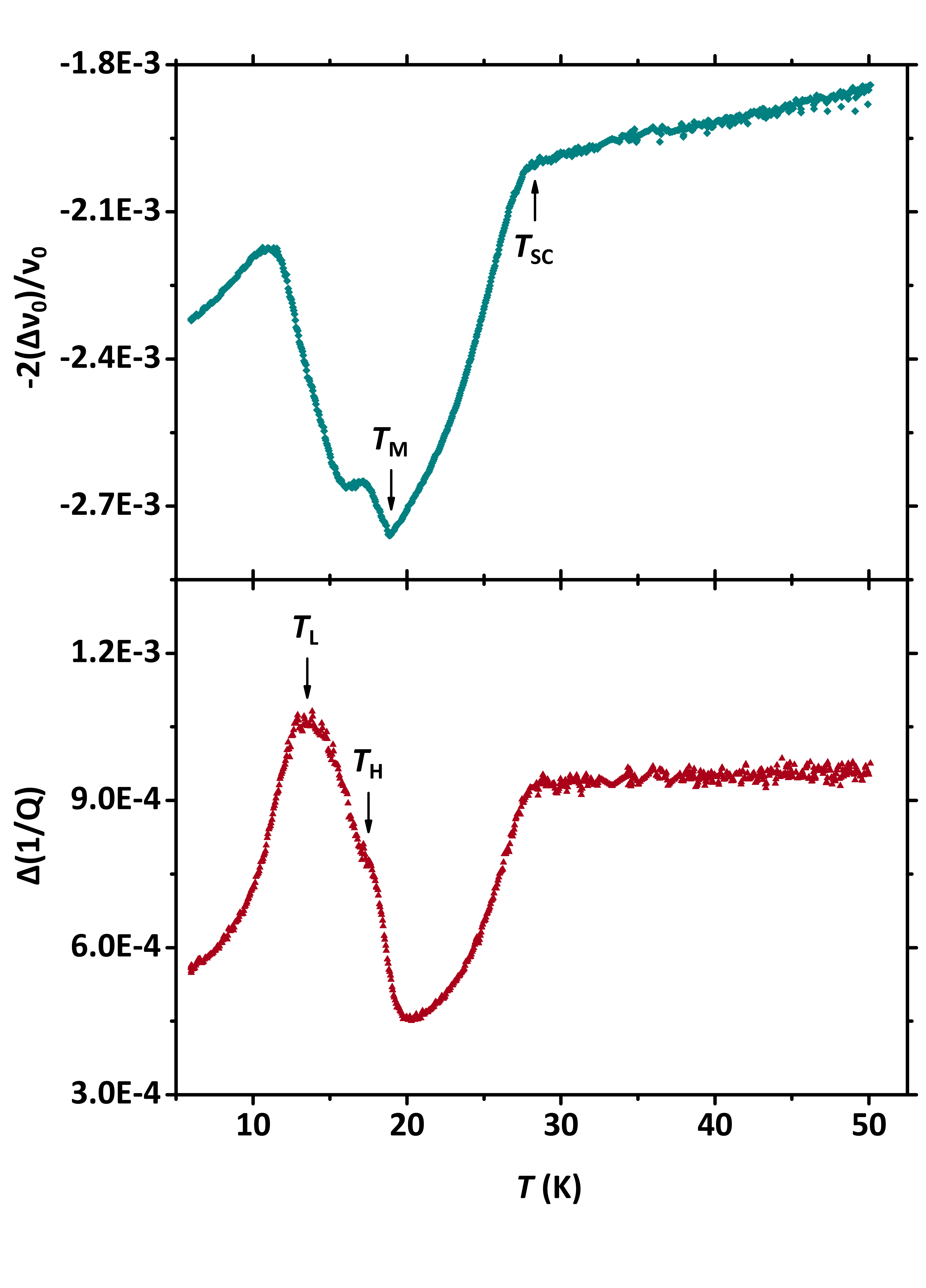}
		\caption{\label{Fig_CPWR} Fractional shifts of the resonance frequency $\nu_0$ (upper panel) and of the quality factor inverse $1/Q$ (lower panel) of the CPWR, due to the presence of the sample under test, at $\nu\simeq7.9$~GHz. According to Eq.~\ref{eq.CPWR}, these quantities are proportional to the high-frequency magnetic susceptibility real and imaginary parts: $\Delta \nu_0/\nu_0 \propto \chi'$ and $\Delta(1/Q) \propto \chi''$.}
	\end{figure}
	The fractional shift of the resonance frequency $\nu$ and of the quality factor $Q$ of the CPWR at $\nu \simeq 7.9$ GHz -- corresponding to the real and imaginary components of $\chi$, respectively, according to Eq.~\ref{eq.CPWR} -- are reported in Fig.~\ref{Fig_CPWR}. $\chi^{\prime}$ shows a temperature dependence qualitatively similar to what is measured at lower frequencies. For $T > T_{SC}$, $\chi^{\prime}$ is negative and it takes non-negligible absolute values. This is due to the classical skin depth $\delta$ being lower than the half-thickness of the sample at this high frequency, thus assuring a shielding effect also in the normal state. For the same reason, the values of $\chi^{\prime\prime}$ above $T_{SC}$ remain significantly different from zero: the theory of electrodynamics of normal metals \cite{landau} gives, e.g., for a slab of thickness of the order of $2\delta$ with $H_{ac}$ parallel to the slab surface, $\chi^{\prime\prime} \simeq |\chi^{\prime}| \simeq 0.4$. According to the same theory, upon cooling below $T_{SC}$, $\chi^{\prime\prime}$ decreases, since the London penetration depth becomes lower and lower than $\delta$. At about 19 K, both $\chi^{\prime}$ and $\chi^{\prime\prime}$ deviate from the behavior expected for a pure superconductor, since ferromagnetic order sets up. Upon further cooling, the two features (a clear maximum at $T_L$ and a shoulder revealing a second maximum at $T_H$) already identified at lower frequencies in $\chi^{\prime\prime}$ emerge, although at different temperatures.
	
	\begin{figure}[t!]
		\centering
		\includegraphics[width=0.48\textwidth]{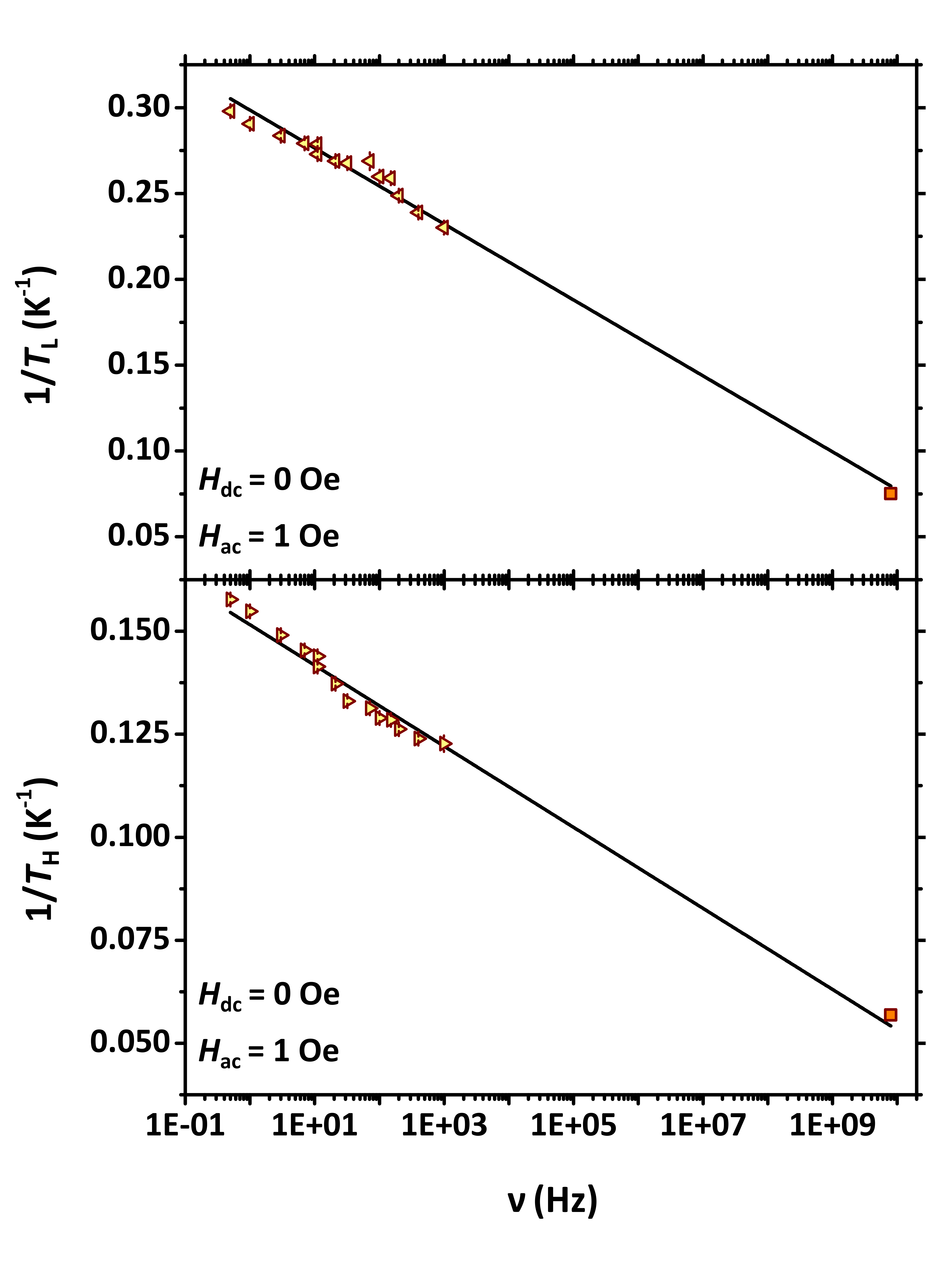}
		\caption{\label{FigNuDepMax} Dependence of the inverse of $T_L$ and $T_H$ on frequency (upper panel and lower panel, respectively). The triangles and the squares refer to ac susceptometry and CPWR-based measurements, respectively. The continuous lines are best-fitting functions based on the activated trend reported in Eq.~\ref{EqFreqDep}.}
	\end{figure}
	The dependence of the characteristic temperature of the two anomalies on frequency is shown in Fig.~\ref{FigNuDepMax}. In the same graphs, we report the temperature position of the anomalies detected by means of standard ac susceptometry and CPWR technique, encompassing $10$ decades for the frequency value. We fit the whole experimental dataset satisfactorily to the activated-like trend
	\begin{equation}\label{EqFreqDep}
		\frac{1}{T_{H,L}} = -\frac{k_{B}}{U_{0}}\ln\left(\frac{\nu}{\nu_{0}}\right).
	\end{equation}
	As a result, we infer the activation energies $U_{0} = 104 \pm 3$ K and $U_{0} = 234 \pm 7$ K for the anomalies at $T_L$ and $T_H$, respectively, with attempt frequencies $\nu_{0} \sim 10^{13}$ -- $10^{15}$ Hz.
	
	\section{Discussion}\label{SectDisc}
	
	Muons couple very effectively with the magnetic phase of the sample, which originates below $T_{M}$ from the long-ranged ordering of Eu$^{2+}$ magnetic moments over the whole sample volume with a large internal magnetic field at the muon site saturating at $\sim 12$ kG, in agreement with previous reports \cite{Guguchia}. Combined with the sharp onset of bulk diamagnetism at $T_{SC}$ observed by $\chi$ measurements, our results confirm the spatial coexistence of superconductivity and magnetism. Values $\sim 0.3$ for the critical exponent $\beta$ are observed commonly for phase transitions on three-dimensional lattices \cite{ChaikinLubensky} while the $T^{3/2}$-like reduction of the magnetization shown in Fig.~\ref{FigMuSRfitNew1} is accounted for by a conventional Bloch-like law characteristic of the excitation of ferromagnetic magnons with quadratic dispersion \cite{Dyson1956,VanVleck1958,Ziman}. Overall, these results point to a conventional three-dimensional ferromagnetic ground state.
	
	Particularly relevant for the following discussion is the dependence of the longitudinal relaxation rate $\lambda_L$ on temperature (see Fig.~\ref{FigMuSRfitNew2}). The observed power-law trend is the expected behaviour for relaxation processes driven by magnon scattering, the precise value of the exponent being dependent on the details of the scattering process being considered \cite{Beeman1968,YaouancDeReotier}. The absence of additional features in the dependence of $\lambda_L$ on temperature below $T_{M}$ indicates that no reorientation transitions take place. Moreover, the low $\lambda_L$ values found in the ordered phase suggest that the movement of magnetic domain walls also does not play an important role. Indeed, a much slower decrease of $\lambda_L$ on cooling within the magnetic phase was associated to the random movement of magnetic domain walls in other Eu based compounds \cite{Eu227}.
	
	It should be remarked that the complex magnetic domain structure characteristic of EuFe$_2$(As$_{1-x}$P$_{x}$)$_2$ is responsible for the generation of superconducting vortices. These domains were extensively studied by magnetic force microscopy on single crystals of the same material in Refs.~\cite{Stolyarov2018sciadv,Ghigo2019PRR}, where it was shown that spontaneous vortex (V) -- antivortex (AV) pairs nucleate locally over a pre-existent striped Meissner structure and evolve in complex arrangements at lower temperatures, decorating Turing-like magnetic domain patterns. Besides the generation of vortices, such magnetic domain structures provide the pinning background where V--AV pairs develop interesting dynamical regimes which, in principle, should contribute to $\lambda_L$ as well with the following functional form \cite{YaouancDeReotier,PrandoNL}
	\begin{equation}\label{EqLambda}
		\lambda_{L}^{V-AV}(T) = 2 \gamma_{\mu}^{2} \Delta^{2} \frac{\tau_{c}(T)}{1 + \omega_{\mu}^{2} \tau_{c}^{2}(T)}.
	\end{equation}
	Here, $\Delta$ is the root-mean-square amplitude of the fluctuations of the local magnetic field, $\tau_{c}$ is the correlation time of the dynamics being probed and $\omega_{\mu} = \gamma_{\mu} B_{\mu}$ is the Larmor angular frequency of the muon. The maximum value of the function in Eq.~\ref{EqLambda} is
	\begin{equation}\label{EqLambdaMax}
		\left.\lambda_{L}^{V-AV}\right|_{max} = \frac{\gamma_{\mu} \Delta^{2}}{B_{\mu}}
	\end{equation}
	and it is taken when $\omega_{\mu} \tau_{c} = 1$. Assuming $\Delta \sim 10$ G as typical value for the fluctuating magnetic field induced by the motion of vortices \cite{Carretta2020RNC} and $B_{\mu} \sim 1$ kG as lower bound for the magnetic field at the muon site, we estimate $\left.\lambda_{L}^{V-AV}\right|_{max} \sim 8.5 \times 10^{-3} \; \mu$s$^{-1}$ as upper-bound value of the maximum relaxation rate. The typical experimental values reported for $\lambda_{L}$ in Fig.~\ref{FigMuSRfitNew2} and determined by the contribution of magnons to the relaxation are systematically higher than this estimate, making it clear that muons can hardly resolve the dynamics associated with V--AV pairs in the investigated EuFe$_{2}$(As$_{0.7}$P$_{0.3}$)$_{2}$ sample.
	
	On the other hand, we showed in Sect.~\ref{SectRes} that our $\chi$ measurements  probe features indicative of activated dynamics at low temperatures. In particular, when measuring $\chi^{\prime\prime}$ at a frequency value $\nu$, a peak is resolved with a maximum at the temperature value $T_{p}$ when the condition
	\begin{equation}\label{EqMaximum}
		2\pi\nu\tau_{c|_{T = T_{p}}} = 1
	\end{equation}
	is matched. If these processes were of magnetic origin and associated with the strong magnetic moments localized on the Eu$^{2+}$ ions, the root-mean-square amplitude of the fluctuations of the local magnetic field would be $\Delta \gg 10$ G. Based on Eqs.~\ref{EqLambda} and \ref{EqLambdaMax} and taking $\tau_c=1/\left(2\pi\nu\right)$ from Fig.~\ref{FigNuDepMax}, an extra peak would clearly appear in $\lambda_L$ upon approaching the resonant condition $\omega_{\mu} \tau_c = 1$ for $\sim 10$ -- $17$ K. However, as mentioned above, we have no indication of contributions to the muon longitudinal relaxation other than the magnon scattering at low temperatures. Accordingly, we argue that the activated trends detected by $\chi^{\prime\prime}$ are directly connected to the dynamics of spontaneous V--AV pairs, suggesting that the depinning of vortices -- possibly of collective origin -- is the relevant underlying physical process \cite{Prando2011,Prando2012,Prando2013}. A similar activated trend was found in another Eu-based pnictide, even though the higher depinning energy barriers indicated much stronger intrinsic pinning in that case \cite{Vlasenko2020sust}. Comparable activation energies were found for the collective pinning regime in other pnictides \cite{Wang2017APL} especially under the effect of strong polarizing magnetic fields \cite{Prando2011,Prando2012} which could be mimicked in the current material by the strong internal fields arising from the ferromagnetic background. Also, the observed activated trends for both anomalies exclude a V--AV dissociation transition of the type predicted by the theory of Kosterlitz and Thouless \cite{Hebard1980prl}, that would rather imply a divergent trend near the transition.
	
	\begin{figure}[t!]
		\centering
		\includegraphics[width=0.47\textwidth]{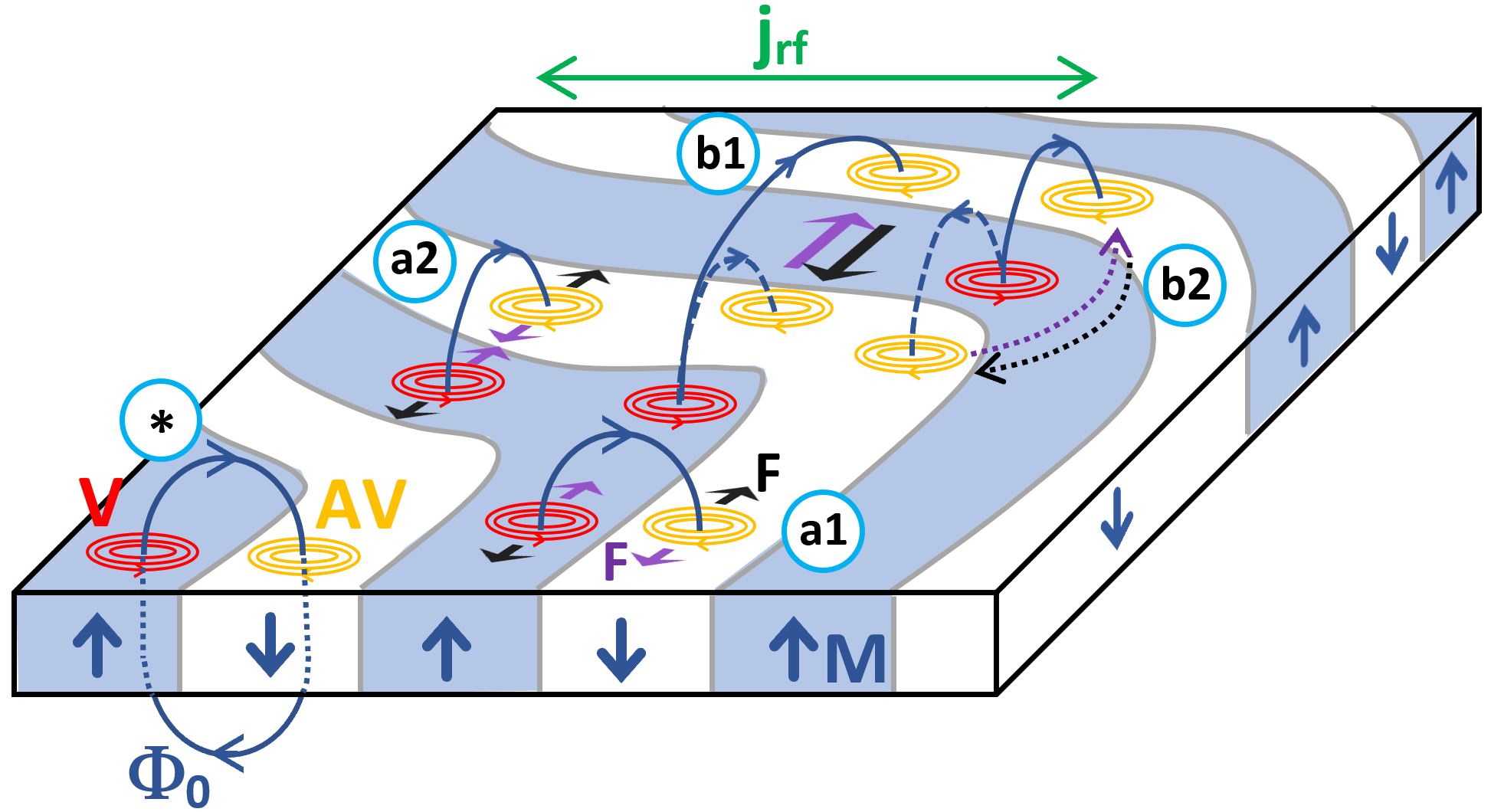}
		\caption{\label{schema} Sketch of possible vortex-antivortex dynamics schemes. The crystal shows magnetic domains (alternate blue and white regions) with out-of-plane magnetization (M, dark blue). A vortex (V, red) - antivortex (AV, orange) pair, spontaneously generated by the magnetic structure, is shown in \textbf{($\ast$)}, coupled by a magnetic flux line loop ($\Phi_0$, dark blue). The green arrow defines the direction of the MW currents, that exert on Vs and AVs Lorentz forces represented as in-plane black and purple arrows (F). In the schemes \textbf{(a1)}-\textbf{(a2)}, Vs and AVs move within single striped magnetic domains, either longitudinally \textbf{(a1)} or transversally \textbf{(a2)}. Although the sketch is not strictly to scale, the dimensions of the vortices and the domains width are effectively comparable. Further schemes represent inter-domain motions. In the scheme \textbf{(b1)} an AV jumps to the next equally-oriented domain, causing the enlargement of the loop. In \textbf{(b2)}, an AV jumps, producing a rotation of the loop without enlargement.}
	\end{figure}
	
	Two distinct peaks in $\chi^{\prime\prime}$ are often resolved in powder samples with low connectivity due to the presence of intra-grain and inter-grains depinning processes, and their assignment is normally made possible based on the markedly different characteristic temperature widths \cite{Gomory1997sust,Prando2012}. However, this scenario is not satisfactory for our current powder sample as we observed two peaks in $\chi^{\prime\prime}$ for single-crystal samples as well \cite{Ghigo2019PRR,Ghigo2020sust}. A plausible explanation of the observed phenomenology comes from the rather complicated geometry of the magnetic domains in which the Vs and AVs are confined. In particular, we suggest a magnetic analogue of the granular origin of the two activated trends which should be associated to ``intra-domain'' and ``inter-domain'' depinning, respectively (see Fig.~\ref{schema}). Intra-domain processes, labelled a$1$ and a$2$, involve the activated motion of V--AV pairs such that the V is confined within a single magnetic domain and the AV is similarly confined to a nearby domain with opposite magnetization. The elongated, meander-like shape of magnetic domains -- whose characteristic width is comparable to the lateral dimensions of Vs and AVs \cite{Stolyarov2018sciadv,Devizorova2019PRL} -- makes the pinning potential markedly anisotropic, so that the motion along the domain or transversal to the domain (a$1$ and a$2$ in Fig.~\ref{schema}, respectively) should be considered as distinct processes, similarly to the case of superconductor/ferromagnet bilayers \cite{Belkin2008prb}. In principle, anisotropic pinning could explain by itself the presence of two peaks in $\chi$. However, we argue that -- due to the random orientation of the ac field with respect to the domains -- this would give rise to a single broad peak in $\chi$ rather than two distinct features. We notice that the intra-domain motions activated by the ac magnetic field are equivalent to a breathing mode of the V--AV loop with small relative displacement between them.
	
	On the other hand, inter-domain depinning processes involve the tunnelling for either the V or the AV between two domains with the same magnetization, similarly to what was observed in artificially nanostructured systems \cite{Lange2005prb}, and we show two possible representative processes belonging to this class in Fig.~\ref{schema} (labelled b$1$ and b$2$). First, we notice that b$1$-like processes affect strongly the dimensions of the V--AV loop and, as such, should be more energetically costly if compared to the breathing modes realized by the intra-domain motions. At the same time, b$2$-like processes realize the rotation of the V--AV loop around one of the two fluxons, thus preserving its dimensions but still implying the annihilation of an AV (V) in a down (up) domain and simultaneous generation of another one in the second adjacent domain with the same magnetization \cite{Lange2005prb}.
	
	Overall, we argue that all inter-domain depinning processes imply a higher energy cost if compared to the intra-domain dynamics. Accordingly, we propose that the low-barrier and high-barrier activated trends detected by $\chi$ measurements should be associated with intra-domain and inter-domain processes, respectively. An open question concerns the experimental capability to resolve the anisotropic intra-domain depinning caused by the domain shape (a$1$ and a$2$ processes), motivating \textit{ad-hoc} designed experiment with single vortex imaging techniques or $\chi$ measurements under different experimental conditions -- e.g., varying amplitude for the ac magnetic field.
	
	\section{Summarizing remarks and conclusions}\label{SectConcl}
	
	Based on measurements of complex magnetic ac susceptibility with frequency spanning over more than ten orders of magnitude, achieved by means of a combination of conventional ac susceptometry and a novel technique based on a microwave coplanar waveguide resonator, we reported on the complex dynamics characterizing the superconducting state of the optimally-doped EuFe$_2$(As$_{1-x}$P$_{x}$)$_2$, where a local coexistence with ferromagnetism is realized. Assisted by the results of muon spin spectroscopy measurements of the same sample, we evidenced the development of two distinct activated processes which we associate to the motion of spontaneously-generated vortices and antivortices within the meander-like pinning potential provided by the Turing pattern of Meissner magnetic domains.
	
	\section*{Acknowledgements}
	This work was partially supported by the Italian Ministry of Education, University and Research (Project PRIN HIBiSCUS, Grant No. 201785KWLE), and partly by a Grant-in-Aid for Scientific Research (A) (17H01141) from the Japan Society for the Promotion of Science (JSPS). This work is based on experiments performed at the Swiss Muon Source S$\mu$S, Paul Scherrer Institute, Villigen, Switzerland. We gratefully acknowledge the Science and Technology Facilities Council (STFC) for access to muon beamtime at the ISIS Neutron and Muon Source (HiFi facility), and also for the provision of sample preparation. We thank Chennan Wang and Adam Berlie for technical support during the experiments at S$\mu$S and ISIS. S. S. thanks Consiglio Nazionale delle Ricerche (CNR) for supporting his visit at ISIS.
	
	\providecommand{\noopsort}[1]{}\providecommand{\singleletter}[1]{#1}%


\begin{thebibliography}{45}%
		\makeatletter
		\providecommand \@ifxundefined [1]{%
			\@ifx{#1\undefined}
		}%
		\providecommand \@ifnum [1]{%
			\ifnum #1\expandafter \@firstoftwo
			\else \expandafter \@secondoftwo
			\fi
		}%
		\providecommand \@ifx [1]{%
			\ifx #1\expandafter \@firstoftwo
			\else \expandafter \@secondoftwo
			\fi
		}%
		\providecommand \natexlab [1]{#1}%
		\providecommand \enquote  [1]{``#1''}%
		\providecommand \bibnamefont  [1]{#1}%
		\providecommand \bibfnamefont [1]{#1}%
		\providecommand \citenamefont [1]{#1}%
		\providecommand \href@noop [0]{\@secondoftwo}%
		\providecommand \href [0]{\begingroup \@sanitize@url \@href}%
		\providecommand \@href[1]{\@@startlink{#1}\@@href}%
		\providecommand \@@href[1]{\endgroup#1\@@endlink}%
		\providecommand \@sanitize@url [0]{\catcode `\\12\catcode `\$12\catcode
			`\&12\catcode `\#12\catcode `\^12\catcode `\_12\catcode `\%12\relax}%
		\providecommand \@@startlink[1]{}%
		\providecommand \@@endlink[0]{}%
		\providecommand \url  [0]{\begingroup\@sanitize@url \@url }%
		\providecommand \@url [1]{\endgroup\@href {#1}{\urlprefix }}%
		\providecommand \urlprefix  [0]{URL }%
		\providecommand \Eprint [0]{\href }%
		\providecommand \doibase [0]{https://doi.org/}%
		\providecommand \selectlanguage [0]{\@gobble}%
		\providecommand \bibinfo  [0]{\@secondoftwo}%
		\providecommand \bibfield  [0]{\@secondoftwo}%
		\providecommand \translation [1]{[#1]}%
		\providecommand \BibitemOpen [0]{}%
		\providecommand \bibitemStop [0]{}%
		\providecommand \bibitemNoStop [0]{.\EOS\space}%
		\providecommand \EOS [0]{\spacefactor3000\relax}%
		\providecommand \BibitemShut  [1]{\csname bibitem#1\endcsname}%
		\let\auto@bib@innerbib\@empty
		%</preamble>
		\bibitem [{\citenamefont {Ren}\ \emph {et~al.}(2009)\citenamefont {Ren},
			\citenamefont {Tao}, \citenamefont {Jiang}, \citenamefont {Feng},
			\citenamefont {Wang}, \citenamefont {Dai}, \citenamefont {Cao},\ and\
			\citenamefont {Xu}}]{Ren2009PRB}%
		\BibitemOpen
		\bibfield  {author} {\bibinfo {author} {\bibfnamefont {Z.}~\bibnamefont
				{Ren}}, \bibinfo {author} {\bibfnamefont {Q.}~\bibnamefont {Tao}}, \bibinfo
			{author} {\bibfnamefont {S.}~\bibnamefont {Jiang}}, \bibinfo {author}
			{\bibfnamefont {C.}~\bibnamefont {Feng}}, \bibinfo {author} {\bibfnamefont
				{C.}~\bibnamefont {Wang}}, \bibinfo {author} {\bibfnamefont {J.}~\bibnamefont
				{Dai}}, \bibinfo {author} {\bibfnamefont {G.}~\bibnamefont {Cao}},\ and\
			\bibinfo {author} {\bibfnamefont {Z.}~\bibnamefont {Xu}},\ }\bibfield
		{title} {\bibinfo {title} {Superconductivity induced by phosphorus doping and
				its coexistence with ferromagnetism in
				{EuFe$_{2}$(As$_{0.7}$P$_{0.3}$)$_{2}$}},\ }\href@noop {} {\bibfield
			{journal} {\bibinfo  {journal} {Phys. Rev. Lett.}\ }\textbf {\bibinfo
				{volume} {102}},\ \bibinfo {pages} {137002} (\bibinfo {year}
			{2009})}\BibitemShut {NoStop}%
		\bibitem [{\citenamefont {Cao}\ \emph {et~al.}(2011)\citenamefont {Cao},
			\citenamefont {Xu}, \citenamefont {Ren}, \citenamefont {Jiang}, \citenamefont
			{Feng},\ and\ \citenamefont {Xu}}]{Cao2011JPCM}%
		\BibitemOpen
		\bibfield  {author} {\bibinfo {author} {\bibfnamefont {G.}~\bibnamefont
				{Cao}}, \bibinfo {author} {\bibfnamefont {S.}~\bibnamefont {Xu}}, \bibinfo
			{author} {\bibfnamefont {Z.}~\bibnamefont {Ren}}, \bibinfo {author}
			{\bibfnamefont {S.}~\bibnamefont {Jiang}}, \bibinfo {author} {\bibfnamefont
				{C.}~\bibnamefont {Feng}},\ and\ \bibinfo {author} {\bibfnamefont
				{Z.}~\bibnamefont {Xu}},\ }\bibfield  {title} {\bibinfo {title}
			{Superconductivity and ferromagnetism in
				{EuFe$_{2}$(As$_{1-x}$P$_{x}$)$_{2}$}},\ }\href@noop {} {\bibfield  {journal}
			{\bibinfo  {journal} {J. Phys. Condens. Matter}\ }\textbf {\bibinfo {volume}
				{23}},\ \bibinfo {pages} {464204} (\bibinfo {year} {2011})}\BibitemShut
		{NoStop}%
		\bibitem [{\citenamefont {Zapf}\ \emph {et~al.}(2013)\citenamefont {Zapf},
			\citenamefont {Jeevan}, \citenamefont {Ivek}, \citenamefont {Pfister},
			\citenamefont {Klingert}, \citenamefont {Jiang}, \citenamefont {Wu},
			\citenamefont {Gegenwart}, \citenamefont {Kremer},\ and\ \citenamefont
			{Dressel}}]{Zapf2013PRL}%
		\BibitemOpen
		\bibfield  {author} {\bibinfo {author} {\bibfnamefont {S.}~\bibnamefont
				{Zapf}}, \bibinfo {author} {\bibfnamefont {H.~S.}\ \bibnamefont {Jeevan}},
			\bibinfo {author} {\bibfnamefont {T.}~\bibnamefont {Ivek}}, \bibinfo {author}
			{\bibfnamefont {F.}~\bibnamefont {Pfister}}, \bibinfo {author} {\bibfnamefont
				{F.}~\bibnamefont {Klingert}}, \bibinfo {author} {\bibfnamefont
				{S.}~\bibnamefont {Jiang}}, \bibinfo {author} {\bibfnamefont
				{D.}~\bibnamefont {Wu}}, \bibinfo {author} {\bibfnamefont {P.}~\bibnamefont
				{Gegenwart}}, \bibinfo {author} {\bibfnamefont {R.~K.}\ \bibnamefont
				{Kremer}},\ and\ \bibinfo {author} {\bibfnamefont {M.}~\bibnamefont
				{Dressel}},\ }\bibfield  {title} {\bibinfo {title}
			{{EuFe$_{2}$(As$_{1-x}$P$_{x}$)$_{2}$}: reentrant spin glass and
				superconductivity},\ }\href@noop {} {\bibfield  {journal} {\bibinfo
				{journal} {Phys. Rev. Lett.}\ }\textbf {\bibinfo {volume} {110}},\ \bibinfo
			{pages} {237002} (\bibinfo {year} {2013})}\BibitemShut {NoStop}%
		\bibitem [{\citenamefont {Miclea}\ \emph {et~al.}(2009)\citenamefont {Miclea},
			\citenamefont {Nicklas}, \citenamefont {Jeevan}, \citenamefont {Kasinathan},
			\citenamefont {Hossain}, \citenamefont {Rosner}, \citenamefont {Gegenwart},
			\citenamefont {Geibel},\ and\ \citenamefont {Steglich}}]{Miclea2009PRB}%
		\BibitemOpen
		\bibfield  {author} {\bibinfo {author} {\bibfnamefont {C.~F.}\ \bibnamefont
				{Miclea}}, \bibinfo {author} {\bibfnamefont {M.}~\bibnamefont {Nicklas}},
			\bibinfo {author} {\bibfnamefont {H.~S.}\ \bibnamefont {Jeevan}}, \bibinfo
			{author} {\bibfnamefont {D.}~\bibnamefont {Kasinathan}}, \bibinfo {author}
			{\bibfnamefont {Z.}~\bibnamefont {Hossain}}, \bibinfo {author} {\bibfnamefont
				{H.}~\bibnamefont {Rosner}}, \bibinfo {author} {\bibfnamefont
				{P.}~\bibnamefont {Gegenwart}}, \bibinfo {author} {\bibfnamefont
				{C.}~\bibnamefont {Geibel}},\ and\ \bibinfo {author} {\bibfnamefont
				{F.}~\bibnamefont {Steglich}},\ }\bibfield  {title} {\bibinfo {title}
			{Evidence for a reentrant superconducting state in {EuFe$_{2}$As$_{2}$} under
				pressure},\ }\href@noop {} {\bibfield  {journal} {\bibinfo  {journal} {Phys.
					Rev. B}\ }\textbf {\bibinfo {volume} {79}},\ \bibinfo {pages} {212509}
			(\bibinfo {year} {2009})}\BibitemShut {NoStop}%
		\bibitem [{\citenamefont {Tokiwa}\ \emph {et~al.}(2012)\citenamefont {Tokiwa},
			\citenamefont {H\"ubner}, \citenamefont {Beck}, \citenamefont {Jeevan},\ and\
			\citenamefont {Gegenwart}}]{Tokiwa2012PRB}%
		\BibitemOpen
		\bibfield  {author} {\bibinfo {author} {\bibfnamefont {Y.}~\bibnamefont
				{Tokiwa}}, \bibinfo {author} {\bibfnamefont {S.-H.}\ \bibnamefont
				{H\"ubner}}, \bibinfo {author} {\bibfnamefont {O.}~\bibnamefont {Beck}},
			\bibinfo {author} {\bibfnamefont {H.~S.}\ \bibnamefont {Jeevan}},\ and\
			\bibinfo {author} {\bibfnamefont {P.}~\bibnamefont {Gegenwart}},\ }\bibfield
		{title} {\bibinfo {title} {Unique phase diagram with narrow superconducting
				dome in {EuFe$_{2}$(As$_{1-x}$P$_{x}$)$_{2}$} due to {Eu}$^{2+}$ local
				magnetic moments},\ }\href@noop {} {\bibfield  {journal} {\bibinfo  {journal}
				{Phys. Rev. B}\ }\textbf {\bibinfo {volume} {86}},\ \bibinfo {pages}
			{220505(R)} (\bibinfo {year} {2012})}\BibitemShut {NoStop}%
		\bibitem [{\citenamefont {Ghigo}\ \emph {et~al.}(2020)\citenamefont {Ghigo},
			\citenamefont {Torsello}, \citenamefont {Gerbaldo}, \citenamefont
			{Gozzelino}, \citenamefont {Pyon}, \citenamefont {Veshchunov}, \citenamefont
			{Tamegai},\ and\ \citenamefont {Cao}}]{Ghigo2020sust}%
		\BibitemOpen
		\bibfield  {author} {\bibinfo {author} {\bibfnamefont {G.}~\bibnamefont
				{Ghigo}}, \bibinfo {author} {\bibfnamefont {D.}~\bibnamefont {Torsello}},
			\bibinfo {author} {\bibfnamefont {R.}~\bibnamefont {Gerbaldo}}, \bibinfo
			{author} {\bibfnamefont {L.}~\bibnamefont {Gozzelino}}, \bibinfo {author}
			{\bibfnamefont {S.}~\bibnamefont {Pyon}}, \bibinfo {author} {\bibfnamefont
				{I.~S.}\ \bibnamefont {Veshchunov}}, \bibinfo {author} {\bibfnamefont
				{T.}~\bibnamefont {Tamegai}},\ and\ \bibinfo {author} {\bibfnamefont {G.-H.}\
				\bibnamefont {Cao}},\ }\bibfield  {title} {\bibinfo {title} {Effects of
				proton irradiation on the magnetic superconductor
				{EuFe$_{2}$(As$_{1-x}$P$_{x}$)$_{2}$}},\ }\href@noop {} {\bibfield  {journal}
			{\bibinfo  {journal} {Supercond. Sci. Technol.}\ }\textbf {\bibinfo {volume}
				{33}},\ \bibinfo {pages} {094011} (\bibinfo {year} {2020})}\BibitemShut
		{NoStop}%
		\bibitem [{\citenamefont {Ghimire}\ \emph {et~al.}(2021)\citenamefont
			{Ghimire}, \citenamefont {Konczykowski}, \citenamefont {Cho}, \citenamefont
			{Tanatar}, \citenamefont {Torsello}, \citenamefont {Veshchunov},
			\citenamefont {Tamegai}, \citenamefont {Ghigo},\ and\ \citenamefont
			{Prozorov}}]{Ghimire2021Materials}%
		\BibitemOpen
		\bibfield  {author} {\bibinfo {author} {\bibfnamefont {S.}~\bibnamefont
				{Ghimire}}, \bibinfo {author} {\bibfnamefont {M.}~\bibnamefont
				{Konczykowski}}, \bibinfo {author} {\bibfnamefont {K.}~\bibnamefont {Cho}},
			\bibinfo {author} {\bibfnamefont {M.~A.}\ \bibnamefont {Tanatar}}, \bibinfo
			{author} {\bibfnamefont {D.}~\bibnamefont {Torsello}}, \bibinfo {author}
			{\bibfnamefont {I.~S.}\ \bibnamefont {Veshchunov}}, \bibinfo {author}
			{\bibfnamefont {T.}~\bibnamefont {Tamegai}}, \bibinfo {author} {\bibfnamefont
				{G.}~\bibnamefont {Ghigo}},\ and\ \bibinfo {author} {\bibfnamefont
				{R.}~\bibnamefont {Prozorov}},\ }\bibfield  {title} {\bibinfo {title} {Effect
				of controlled artificial disorder on the magnetic properties of
				{EuFe$_{2}$(As$_{1-x}$P$_{x}$)$_{2}$} ferromagnetic superconductor},\
		}\href@noop {} {\bibfield  {journal} {\bibinfo  {journal} {Materials}\
			}\textbf {\bibinfo {volume} {14}},\ \bibinfo {pages} {3267} (\bibinfo {year}
			{2021})}\BibitemShut {NoStop}%
		\bibitem [{\citenamefont {Liu}\ \emph {et~al.}(2016)\citenamefont {Liu},
			\citenamefont {Liu}, \citenamefont {Tang}, \citenamefont {Jiang},
			\citenamefont {Wang}, \citenamefont {Ablimit}, \citenamefont {Jiao},
			\citenamefont {Tao}, \citenamefont {Feng}, \citenamefont {Xu},\ and\
			\citenamefont {Cao}}]{Liu2016PRB}%
		\BibitemOpen
		\bibfield  {author} {\bibinfo {author} {\bibfnamefont {Y.}~\bibnamefont
				{Liu}}, \bibinfo {author} {\bibfnamefont {Y.-B.}\ \bibnamefont {Liu}},
			\bibinfo {author} {\bibfnamefont {Z.-T.}\ \bibnamefont {Tang}}, \bibinfo
			{author} {\bibfnamefont {H.}~\bibnamefont {Jiang}}, \bibinfo {author}
			{\bibfnamefont {Z.-C.}\ \bibnamefont {Wang}}, \bibinfo {author}
			{\bibfnamefont {A.}~\bibnamefont {Ablimit}}, \bibinfo {author} {\bibfnamefont
				{W.-H.}\ \bibnamefont {Jiao}}, \bibinfo {author} {\bibfnamefont
				{Q.}~\bibnamefont {Tao}}, \bibinfo {author} {\bibfnamefont {C.-M.}\
				\bibnamefont {Feng}}, \bibinfo {author} {\bibfnamefont {Z.-A.}\ \bibnamefont
				{Xu}},\ and\ \bibinfo {author} {\bibfnamefont {G.-H.}\ \bibnamefont {Cao}},\
		}\bibfield  {title} {\bibinfo {title} {Superconductivity and ferromagnetism
				in hole-doped {RbEuFe$_{4}$As$_{4}$}},\ }\href@noop {} {\bibfield  {journal}
			{\bibinfo  {journal} {Phys. Rev. B}\ }\textbf {\bibinfo {volume} {93}},\
			\bibinfo {pages} {214503} (\bibinfo {year} {2016})}\BibitemShut {NoStop}%
		\bibitem [{\citenamefont {Smylie}\ \emph {et~al.}(2019)\citenamefont {Smylie},
			\citenamefont {Koshelev}, \citenamefont {Willa}, \citenamefont {Willa},
			\citenamefont {Kwok}, \citenamefont {Bao}, \citenamefont {Chung},
			\citenamefont {Kanatzidis}, \citenamefont {Singleton}, \citenamefont
			{Balakirev}, \citenamefont {Hebbeker}, \citenamefont {Niraula}, \citenamefont
			{Bokari}, \citenamefont {Kayani},\ and\ \citenamefont
			{Welp}}]{Smylie2019PRB}%
		\BibitemOpen
		\bibfield  {author} {\bibinfo {author} {\bibfnamefont {M.~P.}\ \bibnamefont
				{Smylie}}, \bibinfo {author} {\bibfnamefont {A.~E.}\ \bibnamefont
				{Koshelev}}, \bibinfo {author} {\bibfnamefont {K.}~\bibnamefont {Willa}},
			\bibinfo {author} {\bibfnamefont {R.}~\bibnamefont {Willa}}, \bibinfo
			{author} {\bibfnamefont {W.-K.}\ \bibnamefont {Kwok}}, \bibinfo {author}
			{\bibfnamefont {J.-K.}\ \bibnamefont {Bao}}, \bibinfo {author} {\bibfnamefont
				{D.~Y.}\ \bibnamefont {Chung}}, \bibinfo {author} {\bibfnamefont {M.~G.}\
				\bibnamefont {Kanatzidis}}, \bibinfo {author} {\bibfnamefont
				{J.}~\bibnamefont {Singleton}}, \bibinfo {author} {\bibfnamefont {F.~F.}\
				\bibnamefont {Balakirev}}, \bibinfo {author} {\bibfnamefont {H.}~\bibnamefont
				{Hebbeker}}, \bibinfo {author} {\bibfnamefont {P.}~\bibnamefont {Niraula}},
			\bibinfo {author} {\bibfnamefont {E.}~\bibnamefont {Bokari}}, \bibinfo
			{author} {\bibfnamefont {A.}~\bibnamefont {Kayani}},\ and\ \bibinfo {author}
			{\bibfnamefont {U.}~\bibnamefont {Welp}},\ }\bibfield  {title} {\bibinfo
			{title} {Anisotropic upper critical field of pristine and proton-irradiated
				single crystals of the magnetically ordered superconductor
				{RbEuFe$_{4}$As$_{4}$}},\ }\href@noop {} {\bibfield  {journal} {\bibinfo
				{journal} {Phys. Rev. B}\ }\textbf {\bibinfo {volume} {100}},\ \bibinfo
			{pages} {054507} (\bibinfo {year} {2019})}\BibitemShut {NoStop}%
		\bibitem [{\citenamefont {Koshelev}(2019)}]{Koshelev2019PRB}%
		\BibitemOpen
		\bibfield  {author} {\bibinfo {author} {\bibfnamefont {A.~E.}\ \bibnamefont
				{Koshelev}},\ }\bibfield  {title} {\bibinfo {title} {Helical structures in
				layered magnetic superconductors due to indirect exchange interactions
				mediated by interlayer tunneling},\ }\href@noop {} {\bibfield  {journal}
			{\bibinfo  {journal} {Phys. Rev. B}\ }\textbf {\bibinfo {volume} {100}},\
			\bibinfo {pages} {224503} (\bibinfo {year} {2019})}\BibitemShut {NoStop}%
		\bibitem [{\citenamefont {Stolyarov}\ \emph {et~al.}(2018)\citenamefont
			{Stolyarov}, \citenamefont {Veshchunov}, \citenamefont {Grebenchuk},
			\citenamefont {Baranov}, \citenamefont {Golovchanskiy}, \citenamefont
			{Shishkin}, \citenamefont {Zhou}, \citenamefont {Shi}, \citenamefont {Xu},
			\citenamefont {Pyon}, \citenamefont {Sun}, \citenamefont {Jiao},
			\citenamefont {Cao}, \citenamefont {Vinnikov}, \citenamefont {Golubov},
			\citenamefont {Tamegai}, \citenamefont {Buzdin},\ and\ \citenamefont
			{Roditchev}}]{Stolyarov2018sciadv}%
		\BibitemOpen
		\bibfield  {author} {\bibinfo {author} {\bibfnamefont {V.~S.}\ \bibnamefont
				{Stolyarov}}, \bibinfo {author} {\bibfnamefont {I.~S.}\ \bibnamefont
				{Veshchunov}}, \bibinfo {author} {\bibfnamefont {S.~Y.}\ \bibnamefont
				{Grebenchuk}}, \bibinfo {author} {\bibfnamefont {D.~S.}\ \bibnamefont
				{Baranov}}, \bibinfo {author} {\bibfnamefont {I.~A.}\ \bibnamefont
				{Golovchanskiy}}, \bibinfo {author} {\bibfnamefont {A.~G.}\ \bibnamefont
				{Shishkin}}, \bibinfo {author} {\bibfnamefont {N.}~\bibnamefont {Zhou}},
			\bibinfo {author} {\bibfnamefont {Z.}~\bibnamefont {Shi}}, \bibinfo {author}
			{\bibfnamefont {X.}~\bibnamefont {Xu}}, \bibinfo {author} {\bibfnamefont
				{S.}~\bibnamefont {Pyon}}, \bibinfo {author} {\bibfnamefont {Y.}~\bibnamefont
				{Sun}}, \bibinfo {author} {\bibfnamefont {W.}~\bibnamefont {Jiao}}, \bibinfo
			{author} {\bibfnamefont {G.-H.}\ \bibnamefont {Cao}}, \bibinfo {author}
			{\bibfnamefont {L.~Y.}\ \bibnamefont {Vinnikov}}, \bibinfo {author}
			{\bibfnamefont {A.~A.}\ \bibnamefont {Golubov}}, \bibinfo {author}
			{\bibfnamefont {T.}~\bibnamefont {Tamegai}}, \bibinfo {author} {\bibfnamefont
				{A.~I.}\ \bibnamefont {Buzdin}},\ and\ \bibinfo {author} {\bibfnamefont
				{D.}~\bibnamefont {Roditchev}},\ }\bibfield  {title} {\bibinfo {title}
			{Domain {Meissner} state and spontaneous vortex-antivortex generation in the
				ferromagnetic superconductor {EuFe$_{2}$(As$_{0.79}$P$_{0.21}$)$_{2}$}},\
		}\href@noop {} {\bibfield  {journal} {\bibinfo  {journal} {Sci. Adv.}\
			}\textbf {\bibinfo {volume} {4}},\ \bibinfo {pages} {eaat1061} (\bibinfo
			{year} {2018})}\BibitemShut {NoStop}%
		\bibitem [{\citenamefont {Vinnikov}\ \emph {et~al.}(2019)\citenamefont
			{Vinnikov}, \citenamefont {Veshchunov}, \citenamefont {Sidel'nikov},
			\citenamefont {Stolyarov}, \citenamefont {Egorov}, \citenamefont {Skryabina},
			\citenamefont {Jiao}, \citenamefont {Cao},\ and\ \citenamefont
			{Tamegai}}]{Vinnikov2019JETP}%
		\BibitemOpen
		\bibfield  {author} {\bibinfo {author} {\bibfnamefont {L.~Y.}\ \bibnamefont
				{Vinnikov}}, \bibinfo {author} {\bibfnamefont {I.~S.}\ \bibnamefont
				{Veshchunov}}, \bibinfo {author} {\bibfnamefont {M.~S.}\ \bibnamefont
				{Sidel'nikov}}, \bibinfo {author} {\bibfnamefont {V.~S.}\ \bibnamefont
				{Stolyarov}}, \bibinfo {author} {\bibfnamefont {S.~V.}\ \bibnamefont
				{Egorov}}, \bibinfo {author} {\bibfnamefont {O.~V.}\ \bibnamefont
				{Skryabina}}, \bibinfo {author} {\bibfnamefont {W.}~\bibnamefont {Jiao}},
			\bibinfo {author} {\bibfnamefont {G.}~\bibnamefont {Cao}},\ and\ \bibinfo
			{author} {\bibfnamefont {T.}~\bibnamefont {Tamegai}},\ }\bibfield  {title}
		{\bibinfo {title} {Direct observation of vortex and {Meissner} domains in a
				ferromagnetic superconductor {EuFe$_{2}$(As$_{0.79}$P$_{0.21}$)$_{2}$} single
				crystal},\ }\href@noop {} {\bibfield  {journal} {\bibinfo  {journal} {JETP
					Lett.}\ }\textbf {\bibinfo {volume} {109}},\ \bibinfo {pages} {521} (\bibinfo
			{year} {2019})}\BibitemShut {NoStop}%
		\bibitem [{\citenamefont {Devizorova}\ \emph {et~al.}(2019)\citenamefont
			{Devizorova}, \citenamefont {Mironov},\ and\ \citenamefont
			{Buzdin}}]{Devizorova2019PRL}%
		\BibitemOpen
		\bibfield  {author} {\bibinfo {author} {\bibfnamefont {Z.}~\bibnamefont
				{Devizorova}}, \bibinfo {author} {\bibfnamefont {S.}~\bibnamefont
				{Mironov}},\ and\ \bibinfo {author} {\bibfnamefont {A.}~\bibnamefont
				{Buzdin}},\ }\bibfield  {title} {\bibinfo {title} {Theory of magnetic domain
				phases in ferromagnetic superconductors},\ }\href@noop {} {\bibfield
			{journal} {\bibinfo  {journal} {Phys. Rev. Lett.}\ }\textbf {\bibinfo
				{volume} {122}},\ \bibinfo {pages} {117002} (\bibinfo {year}
			{2019})}\BibitemShut {NoStop}%
		\bibitem [{\citenamefont {Xiao}\ \emph {et~al.}(2009)\citenamefont {Xiao},
			\citenamefont {Su}, \citenamefont {Meven}, \citenamefont {Mittal},
			\citenamefont {Kumar}, \citenamefont {Chatterji}, \citenamefont {Price},
			\citenamefont {Persson}, \citenamefont {Kumar}, \citenamefont {Dhar},
			\citenamefont {Thamizhavel},\ and\ \citenamefont {Br\"uckel}}]{Xiao2009PRB}%
		\BibitemOpen
		\bibfield  {author} {\bibinfo {author} {\bibfnamefont {Y.}~\bibnamefont
				{Xiao}}, \bibinfo {author} {\bibfnamefont {Y.}~\bibnamefont {Su}}, \bibinfo
			{author} {\bibfnamefont {M.}~\bibnamefont {Meven}}, \bibinfo {author}
			{\bibfnamefont {R.}~\bibnamefont {Mittal}}, \bibinfo {author} {\bibfnamefont
				{C.~M.~N.}\ \bibnamefont {Kumar}}, \bibinfo {author} {\bibfnamefont
				{T.}~\bibnamefont {Chatterji}}, \bibinfo {author} {\bibfnamefont
				{S.}~\bibnamefont {Price}}, \bibinfo {author} {\bibfnamefont
				{J.}~\bibnamefont {Persson}}, \bibinfo {author} {\bibfnamefont
				{N.}~\bibnamefont {Kumar}}, \bibinfo {author} {\bibfnamefont {S.~K.}\
				\bibnamefont {Dhar}}, \bibinfo {author} {\bibfnamefont {A.}~\bibnamefont
				{Thamizhavel}},\ and\ \bibinfo {author} {\bibfnamefont {T.}~\bibnamefont
				{Br\"uckel}},\ }\bibfield  {title} {\bibinfo {title} {Magnetic structure of
				{EuFe$_2$As$_2$} determined by single-crystal neutron diffraction},\
		}\href@noop {} {\bibfield  {journal} {\bibinfo  {journal} {Phys. Rev. B}\
			}\textbf {\bibinfo {volume} {80}},\ \bibinfo {pages} {174424} (\bibinfo
			{year} {2009})}\BibitemShut {NoStop}%
		\bibitem [{\citenamefont {Zapf}\ \emph {et~al.}(2011)\citenamefont {Zapf},
			\citenamefont {Wu}, \citenamefont {Bogani}, \citenamefont {Jeevan},
			\citenamefont {Gegenwart},\ and\ \citenamefont {Dressel}}]{Zapf2011PRB}%
		\BibitemOpen
		\bibfield  {author} {\bibinfo {author} {\bibfnamefont {S.}~\bibnamefont
				{Zapf}}, \bibinfo {author} {\bibfnamefont {D.}~\bibnamefont {Wu}}, \bibinfo
			{author} {\bibfnamefont {L.}~\bibnamefont {Bogani}}, \bibinfo {author}
			{\bibfnamefont {H.~S.}\ \bibnamefont {Jeevan}}, \bibinfo {author}
			{\bibfnamefont {P.}~\bibnamefont {Gegenwart}},\ and\ \bibinfo {author}
			{\bibfnamefont {M.}~\bibnamefont {Dressel}},\ }\bibfield  {title} {\bibinfo
			{title} {Varying {Eu}${}^{2+}$ magnetic order by chemical pressure in
				{EuFe$_{2}$(As$_{1-x}$P$_{x}$)$_{2}$}},\ }\href@noop {} {\bibfield  {journal}
			{\bibinfo  {journal} {Phys. Rev. B}\ }\textbf {\bibinfo {volume} {84}},\
			\bibinfo {pages} {140503} (\bibinfo {year} {2011})}\BibitemShut {NoStop}%
		\bibitem [{\citenamefont {Nandi}\ \emph
			{et~al.}(2014{\natexlab{a}})\citenamefont {Nandi}, \citenamefont {Jin},
			\citenamefont {Xiao}, \citenamefont {Su}, \citenamefont {Price},
			\citenamefont {Schmidt}, \citenamefont {Schmalzl}, \citenamefont {Chatterji},
			\citenamefont {Jeevan}, \citenamefont {Gegenwart},\ and\ \citenamefont
			{Br{\"u}ckel}}]{Nandi2014prb2}%
		\BibitemOpen
		\bibfield  {author} {\bibinfo {author} {\bibfnamefont {S.}~\bibnamefont
				{Nandi}}, \bibinfo {author} {\bibfnamefont {W.~T.}\ \bibnamefont {Jin}},
			\bibinfo {author} {\bibfnamefont {Y.}~\bibnamefont {Xiao}}, \bibinfo {author}
			{\bibfnamefont {Y.}~\bibnamefont {Su}}, \bibinfo {author} {\bibfnamefont
				{S.}~\bibnamefont {Price}}, \bibinfo {author} {\bibfnamefont
				{W.}~\bibnamefont {Schmidt}}, \bibinfo {author} {\bibfnamefont
				{K.}~\bibnamefont {Schmalzl}}, \bibinfo {author} {\bibfnamefont
				{T.}~\bibnamefont {Chatterji}}, \bibinfo {author} {\bibfnamefont {H.~S.}\
				\bibnamefont {Jeevan}}, \bibinfo {author} {\bibfnamefont {P.}~\bibnamefont
				{Gegenwart}},\ and\ \bibinfo {author} {\bibfnamefont {T.}~\bibnamefont
				{Br{\"u}ckel}},\ }\bibfield  {title} {\bibinfo {title} {Magnetic structure of
				the {Eu$^{2+}$} moments in superconducting
				{EuFe$_{2}$(As$_{1-x}$P$_{x}$)$_{2}$} with $x = 0.19$},\ }\href@noop {}
		{\bibfield  {journal} {\bibinfo  {journal} {Phys. Rev. B}\ }\textbf {\bibinfo
				{volume} {90}},\ \bibinfo {pages} {094407} (\bibinfo {year}
			{2014}{\natexlab{a}})}\BibitemShut {NoStop}%
		\bibitem [{\citenamefont {Nandi}\ \emph
			{et~al.}(2014{\natexlab{b}})\citenamefont {Nandi}, \citenamefont {Jin},
			\citenamefont {Xiao}, \citenamefont {Su}, \citenamefont {Price},
			\citenamefont {Shukla}, \citenamefont {Strempfer}, \citenamefont {Jeevan},
			\citenamefont {Gegenwart},\ and\ \citenamefont {Br{\"u}ckel}}]{Nandi2014PRB}%
		\BibitemOpen
		\bibfield  {author} {\bibinfo {author} {\bibfnamefont {S.}~\bibnamefont
				{Nandi}}, \bibinfo {author} {\bibfnamefont {W.~T.}\ \bibnamefont {Jin}},
			\bibinfo {author} {\bibfnamefont {Y.}~\bibnamefont {Xiao}}, \bibinfo {author}
			{\bibfnamefont {Y.}~\bibnamefont {Su}}, \bibinfo {author} {\bibfnamefont
				{S.}~\bibnamefont {Price}}, \bibinfo {author} {\bibfnamefont {D.~K.}\
				\bibnamefont {Shukla}}, \bibinfo {author} {\bibfnamefont {J.}~\bibnamefont
				{Strempfer}}, \bibinfo {author} {\bibfnamefont {H.~S.}\ \bibnamefont
				{Jeevan}}, \bibinfo {author} {\bibfnamefont {P.}~\bibnamefont {Gegenwart}},\
			and\ \bibinfo {author} {\bibfnamefont {T.}~\bibnamefont {Br{\"u}ckel}},\
		}\bibfield  {title} {\bibinfo {title} {Coexistence of superconductivity and
				ferromagnetism in {P}-doped {EuFe$_2$As$_2$}},\ }\href@noop {} {\bibfield
			{journal} {\bibinfo  {journal} {Phys. Rev. B}\ }\textbf {\bibinfo {volume}
				{89}},\ \bibinfo {pages} {014512} (\bibinfo {year}
			{2014}{\natexlab{b}})}\BibitemShut {NoStop}%
		\bibitem [{\citenamefont {Zapf}\ and\ \citenamefont
			{Dressel}(2016)}]{Zapf2016RPP}%
		\BibitemOpen
		\bibfield  {author} {\bibinfo {author} {\bibfnamefont {S.}~\bibnamefont
				{Zapf}}\ and\ \bibinfo {author} {\bibfnamefont {M.}~\bibnamefont {Dressel}},\
		}\bibfield  {title} {\bibinfo {title} {Europium-based iron pnictides: a
				unique laboratory for magnetism, superconductivity and structural effects},\
		}\href@noop {} {\bibfield  {journal} {\bibinfo  {journal} {Rep. Prog. Phys.}\
			}\textbf {\bibinfo {volume} {80}},\ \bibinfo {pages} {016501} (\bibinfo
			{year} {2016})}\BibitemShut {NoStop}%
		\bibitem [{\citenamefont {Ghigo}\ \emph {et~al.}(2019)\citenamefont {Ghigo},
			\citenamefont {Torsello}, \citenamefont {Gozzelino}, \citenamefont {Tamegai},
			\citenamefont {Veshchunov}, \citenamefont {Pyon}, \citenamefont {Jiao},
			\citenamefont {Cao}, \citenamefont {Grebenchuk}, \citenamefont
			{Golovchanskiy}, \citenamefont {Stolyarov},\ and\ \citenamefont
			{Roditchev}}]{Ghigo2019PRR}%
		\BibitemOpen
		\bibfield  {author} {\bibinfo {author} {\bibfnamefont {G.}~\bibnamefont
				{Ghigo}}, \bibinfo {author} {\bibfnamefont {D.}~\bibnamefont {Torsello}},
			\bibinfo {author} {\bibfnamefont {L.}~\bibnamefont {Gozzelino}}, \bibinfo
			{author} {\bibfnamefont {T.}~\bibnamefont {Tamegai}}, \bibinfo {author}
			{\bibfnamefont {I.~S.}\ \bibnamefont {Veshchunov}}, \bibinfo {author}
			{\bibfnamefont {S.}~\bibnamefont {Pyon}}, \bibinfo {author} {\bibfnamefont
				{W.}~\bibnamefont {Jiao}}, \bibinfo {author} {\bibfnamefont {G.-H.}\
				\bibnamefont {Cao}}, \bibinfo {author} {\bibfnamefont {S.~Y.}\ \bibnamefont
				{Grebenchuk}}, \bibinfo {author} {\bibfnamefont {I.~A.}\ \bibnamefont
				{Golovchanskiy}}, \bibinfo {author} {\bibfnamefont {V.~S.}\ \bibnamefont
				{Stolyarov}},\ and\ \bibinfo {author} {\bibfnamefont {D.}~\bibnamefont
				{Roditchev}},\ }\bibfield  {title} {\bibinfo {title} {Microwave analysis of
				the interplay between magnetism and superconductivity in
				{EuFe$_{2}$(As$_{1-x}$P$_{x}$)$_{2}$} single crystals},\ }\href@noop {}
		{\bibfield  {journal} {\bibinfo  {journal} {Phys. Rev. Research}\ }\textbf
			{\bibinfo {volume} {1}},\ \bibinfo {pages} {033110} (\bibinfo {year}
			{2019})}\BibitemShut {NoStop}%
		\bibitem [{\citenamefont {Sanna}(2019)}]{Muons}%
		\BibitemOpen
		\bibfield  {author} {\bibinfo {author} {\bibfnamefont {S.}~\bibnamefont
				{Sanna}},\ }\bibfield  {title} {\bibinfo {title} {Longitudinal relaxation
				measurements in the ferromagnetic phase of the superconducting state of
				eu-based pnictides},\ }\bibfield  {journal} {\bibinfo  {journal} {STFC ISIS
				Neutron and Muon Source}\ }\href {https://doi.org/10.5286/ISIS.E.RB1920606}
		{10.5286/ISIS.E.RB1920606} (\bibinfo {year} {2019})\BibitemShut {NoStop}%
		\bibitem [{\citenamefont {Blundell}(1999)}]{Blundell}%
		\BibitemOpen
		\bibfield  {author} {\bibinfo {author} {\bibfnamefont {S.~J.}\ \bibnamefont
				{Blundell}},\ }\bibfield  {title} {\bibinfo {title} {Spin-polarized muons in
				condensed matter physics},\ }\href@noop {} {\bibfield  {journal} {\bibinfo
				{journal} {Contemp. Phys.}\ }\textbf {\bibinfo {volume} {40}},\ \bibinfo
			{pages} {175} (\bibinfo {year} {1999})}\BibitemShut {NoStop}%
		\bibitem [{\citenamefont {Yaouanc}\ and\ \citenamefont {Dalmas~de
				R\'{e}otier}(2011)}]{YaouancDeReotier}%
		\BibitemOpen
		\bibfield  {author} {\bibinfo {author} {\bibfnamefont {A.}~\bibnamefont
				{Yaouanc}}\ and\ \bibinfo {author} {\bibfnamefont {P.}~\bibnamefont
				{Dalmas~de R\'{e}otier}},\ }\href@noop {} {\emph {\bibinfo {title} {Muon spin
					rotation, relaxation, and resonance. {Applications} to condensed matter}}}\
		(\bibinfo  {publisher} {Oxford University Press},\ \bibinfo {year}
		{2011})\BibitemShut {NoStop}%
		\bibitem [{\citenamefont {Torsello}\ \emph {et~al.}(2019)\citenamefont
			{Torsello}, \citenamefont {Ummarino}, \citenamefont {Gozzelino},
			\citenamefont {Tamegai},\ and\ \citenamefont {Ghigo}}]{Torsello2019PRB}%
		\BibitemOpen
		\bibfield  {author} {\bibinfo {author} {\bibfnamefont {D.}~\bibnamefont
				{Torsello}}, \bibinfo {author} {\bibfnamefont {G.~A.}\ \bibnamefont
				{Ummarino}}, \bibinfo {author} {\bibfnamefont {L.}~\bibnamefont {Gozzelino}},
			\bibinfo {author} {\bibfnamefont {T.}~\bibnamefont {Tamegai}},\ and\ \bibinfo
			{author} {\bibfnamefont {G.}~\bibnamefont {Ghigo}},\ }\bibfield  {title}
		{\bibinfo {title} {Comprehensive {Eliashberg} analysis of microwave
				conductivity and penetration depth of {K-}, {Co-}, and {P-}substituted
				{BaFe$_{2}$As$_{2}$}},\ }\href@noop {} {\bibfield  {journal} {\bibinfo
				{journal} {Phys. Rev. B}\ }\textbf {\bibinfo {volume} {99}},\ \bibinfo
			{pages} {134518} (\bibinfo {year} {2019})}\BibitemShut {NoStop}%
		\bibitem [{\citenamefont {Torsello}\ \emph {et~al.}(2020)\citenamefont
			{Torsello}, \citenamefont {Ummarino}, \citenamefont {Bekaert}, \citenamefont
			{Gozzelino}, \citenamefont {Gerbaldo}, \citenamefont {Tanatar}, \citenamefont
			{Canfield}, \citenamefont {Prozorov},\ and\ \citenamefont
			{Ghigo}}]{Torsello2020PRAppl}%
		\BibitemOpen
		\bibfield  {author} {\bibinfo {author} {\bibfnamefont {D.}~\bibnamefont
				{Torsello}}, \bibinfo {author} {\bibfnamefont {G.}~\bibnamefont {Ummarino}},
			\bibinfo {author} {\bibfnamefont {J.}~\bibnamefont {Bekaert}}, \bibinfo
			{author} {\bibfnamefont {L.}~\bibnamefont {Gozzelino}}, \bibinfo {author}
			{\bibfnamefont {R.}~\bibnamefont {Gerbaldo}}, \bibinfo {author}
			{\bibfnamefont {M.}~\bibnamefont {Tanatar}}, \bibinfo {author} {\bibfnamefont
				{P.}~\bibnamefont {Canfield}}, \bibinfo {author} {\bibfnamefont
				{R.}~\bibnamefont {Prozorov}},\ and\ \bibinfo {author} {\bibfnamefont
				{G.}~\bibnamefont {Ghigo}},\ }\bibfield  {title} {\bibinfo {title} {Tuning
				the intrinsic anisotropy with disorder in the {CaKFe$_{4}$As$_{4}$}
				superconductor},\ }\href@noop {} {\bibfield  {journal} {\bibinfo  {journal}
				{Phys. Rev. Applied}\ }\textbf {\bibinfo {volume} {13}},\ \bibinfo {pages}
			{064046} (\bibinfo {year} {2020})}\BibitemShut {NoStop}%
		\bibitem [{\citenamefont {Ghigo}\ \emph {et~al.}(2017)\citenamefont {Ghigo},
			\citenamefont {Ummarino}, \citenamefont {Gozzelino},\ and\ \citenamefont
			{Tamegai}}]{Ghigo2017PRB}%
		\BibitemOpen
		\bibfield  {author} {\bibinfo {author} {\bibfnamefont {G.}~\bibnamefont
				{Ghigo}}, \bibinfo {author} {\bibfnamefont {G.~A.}\ \bibnamefont {Ummarino}},
			\bibinfo {author} {\bibfnamefont {L.}~\bibnamefont {Gozzelino}},\ and\
			\bibinfo {author} {\bibfnamefont {T.}~\bibnamefont {Tamegai}},\ }\bibfield
		{title} {\bibinfo {title} {Penetration depth of {Ba$_{1-x}$K$_x$Fe$_2$As$_2$}
				single crystals explained within a multiband {Eliashberg} $s \pm$ approach},\
		}\href@noop {} {\bibfield  {journal} {\bibinfo  {journal} {Phys. Rev. B}\
			}\textbf {\bibinfo {volume} {96}},\ \bibinfo {pages} {014501} (\bibinfo
			{year} {2017})}\BibitemShut {NoStop}%
		\bibitem [{\citenamefont {Ghigo}\ and\ \citenamefont {Torsello}(ress)}]{libro}%
		\BibitemOpen
		\bibfield  {author} {\bibinfo {author} {\bibfnamefont {G.}~\bibnamefont
				{Ghigo}}\ and\ \bibinfo {author} {\bibfnamefont {D.}~\bibnamefont
				{Torsello}},\ }\href@noop {} {\emph {\bibinfo {title} {Microwave analysis of
					unconventional superconductors with coplanar-resonator techniques}}}\
		(\bibinfo  {publisher} {Springer Nature},\ \bibinfo {year} {in
			press})\BibitemShut {NoStop}%
		\bibitem [{\citenamefont {Prando}\ \emph
			{et~al.}(2020{\natexlab{a}})\citenamefont {Prando}, \citenamefont {Telang},
			\citenamefont {Wilson}, \citenamefont {Graf},\ and\ \citenamefont
			{Singh}}]{Eu227}%
		\BibitemOpen
		\bibfield  {author} {\bibinfo {author} {\bibfnamefont {G.}~\bibnamefont
				{Prando}}, \bibinfo {author} {\bibfnamefont {P.}~\bibnamefont {Telang}},
			\bibinfo {author} {\bibfnamefont {S.~D.}\ \bibnamefont {Wilson}}, \bibinfo
			{author} {\bibfnamefont {M.~J.}\ \bibnamefont {Graf}},\ and\ \bibinfo
			{author} {\bibfnamefont {S.}~\bibnamefont {Singh}},\ }\bibfield  {title}
		{\bibinfo {title} {Monopole-limited nucleation of magnetism in
				{Eu$_{2}$Ir$_{2}$O$_{7}$}},\ }\href@noop {} {\bibfield  {journal} {\bibinfo
				{journal} {Phys. Rev. B}\ }\textbf {\bibinfo {volume} {101}},\ \bibinfo
			{pages} {174435} (\bibinfo {year} {2020}{\natexlab{a}})}\BibitemShut
		{NoStop}%
		\bibitem [{\citenamefont {Landau}\ \emph {et~al.}(2013)\citenamefont {Landau},
			\citenamefont {Bell}, \citenamefont {Kearsley}, \citenamefont {Pitaevskii},
			\citenamefont {Lifshitz},\ and\ \citenamefont {Sykes}}]{landau}%
		\BibitemOpen
		\bibfield  {author} {\bibinfo {author} {\bibfnamefont {L.~D.}\ \bibnamefont
				{Landau}}, \bibinfo {author} {\bibfnamefont {J.~S.}\ \bibnamefont {Bell}},
			\bibinfo {author} {\bibfnamefont {M.~J.}\ \bibnamefont {Kearsley}}, \bibinfo
			{author} {\bibfnamefont {L.~P.}\ \bibnamefont {Pitaevskii}}, \bibinfo
			{author} {\bibfnamefont {E.~M.}\ \bibnamefont {Lifshitz}},\ and\ \bibinfo
			{author} {\bibfnamefont {J.~B.}\ \bibnamefont {Sykes}},\ }\href@noop {}
		{\emph {\bibinfo {title} {Electrodynamics of continuous media}}}\ (\bibinfo
		{publisher} {Elsevier},\ \bibinfo {year} {2013})\BibitemShut {NoStop}%
		\bibitem [{\citenamefont {Guguchia}\ \emph {et~al.}(2013)\citenamefont
			{Guguchia}, \citenamefont {Shengelaya}, \citenamefont {Maisuradze},
			\citenamefont {Howald}, \citenamefont {Bukowski}, \citenamefont {Chikovani},
			\citenamefont {Luetkens}, \citenamefont {Katrych}, \citenamefont
			{Karpinski},\ and\ \citenamefont {Keller}}]{Guguchia}%
		\BibitemOpen
		\bibfield  {author} {\bibinfo {author} {\bibfnamefont {Z.}~\bibnamefont
				{Guguchia}}, \bibinfo {author} {\bibfnamefont {A.}~\bibnamefont
				{Shengelaya}}, \bibinfo {author} {\bibfnamefont {A.}~\bibnamefont
				{Maisuradze}}, \bibinfo {author} {\bibfnamefont {L.}~\bibnamefont {Howald}},
			\bibinfo {author} {\bibfnamefont {Z.}~\bibnamefont {Bukowski}}, \bibinfo
			{author} {\bibfnamefont {M.}~\bibnamefont {Chikovani}}, \bibinfo {author}
			{\bibfnamefont {H.}~\bibnamefont {Luetkens}}, \bibinfo {author}
			{\bibfnamefont {S.}~\bibnamefont {Katrych}}, \bibinfo {author} {\bibfnamefont
				{J.}~\bibnamefont {Karpinski}},\ and\ \bibinfo {author} {\bibfnamefont
				{H.}~\bibnamefont {Keller}},\ }\bibfield  {title} {\bibinfo {title}
			{Muon-spin rotation and magnetization studies of chemical and hydrostatic
				pressure effects in {EuFe$_{2}$(As$_{1-x}$P$_{x}$)$_{2}$}},\ }\href@noop {}
		{\bibfield  {journal} {\bibinfo  {journal} {J. Supercond. Nov. Magn.}\
			}\textbf {\bibinfo {volume} {26}},\ \bibinfo {pages} {285} (\bibinfo {year}
			{2013})}\BibitemShut {NoStop}%
		\bibitem [{\citenamefont {Chaikin}\ and\ \citenamefont
			{Lubensky}(1995)}]{ChaikinLubensky}%
		\BibitemOpen
		\bibfield  {author} {\bibinfo {author} {\bibfnamefont {P.~M.}\ \bibnamefont
				{Chaikin}}\ and\ \bibinfo {author} {\bibfnamefont {T.~C.}\ \bibnamefont
				{Lubensky}},\ }\href@noop {} {\emph {\bibinfo {title} {Principles of
					condensed matter physics}}}\ (\bibinfo  {publisher} {Cambridge University
			Press},\ \bibinfo {year} {1995})\BibitemShut {NoStop}%
		\bibitem [{\citenamefont {Dyson}(1956)}]{Dyson1956}%
		\BibitemOpen
		\bibfield  {author} {\bibinfo {author} {\bibfnamefont {F.~J.}\ \bibnamefont
				{Dyson}},\ }\bibfield  {title} {\bibinfo {title} {General theory of spin-wave
				interactions},\ }\href@noop {} {\bibfield  {journal} {\bibinfo  {journal}
				{Phys. Rev.}\ }\textbf {\bibinfo {volume} {102}},\ \bibinfo {pages} {1217}
			(\bibinfo {year} {1956})}\BibitemShut {NoStop}%
		\bibitem [{\citenamefont {Van~Kranendonk}\ and\ \citenamefont
			{Van~Vleck}(1958)}]{VanVleck1958}%
		\BibitemOpen
		\bibfield  {author} {\bibinfo {author} {\bibfnamefont {J.}~\bibnamefont
				{Van~Kranendonk}}\ and\ \bibinfo {author} {\bibfnamefont {J.~H.}\
				\bibnamefont {Van~Vleck}},\ }\bibfield  {title} {\bibinfo {title} {Spin
				waves},\ }\href@noop {} {\bibfield  {journal} {\bibinfo  {journal} {Rev. Mod.
					Phys.}\ }\textbf {\bibinfo {volume} {30}},\ \bibinfo {pages} {1} (\bibinfo
			{year} {1958})}\BibitemShut {NoStop}%
		\bibitem [{\citenamefont {Ziman}(1972)}]{Ziman}%
		\BibitemOpen
		\bibfield  {author} {\bibinfo {author} {\bibfnamefont {J.~M.}\ \bibnamefont
				{Ziman}},\ }\href@noop {} {\emph {\bibinfo {title} {Principles of the theory
					of solids}}}\ (\bibinfo  {publisher} {Cambridge University Press},\ \bibinfo
		{year} {1972})\BibitemShut {NoStop}%
		\bibitem [{\citenamefont {Beeman}\ and\ \citenamefont
			{Pincus}(1968)}]{Beeman1968}%
		\BibitemOpen
		\bibfield  {author} {\bibinfo {author} {\bibfnamefont {D.}~\bibnamefont
				{Beeman}}\ and\ \bibinfo {author} {\bibfnamefont {P.}~\bibnamefont
				{Pincus}},\ }\bibfield  {title} {\bibinfo {title} {Nuclear spin-lattice
				relaxation in magnetic insulators},\ }\href@noop {} {\bibfield  {journal}
			{\bibinfo  {journal} {Phys. Rev.}\ }\textbf {\bibinfo {volume} {166}},\
			\bibinfo {pages} {359} (\bibinfo {year} {1968})}\BibitemShut {NoStop}%
		\bibitem [{\citenamefont {Prando}\ \emph
			{et~al.}(2020{\natexlab{b}})\citenamefont {Prando}, \citenamefont {Perego},
			\citenamefont {Negroni}, \citenamefont {Ricc\`o}, \citenamefont {Bracco},
			\citenamefont {Comotti}, \citenamefont {Sozzani},\ and\ \citenamefont
			{Carretta}}]{PrandoNL}%
		\BibitemOpen
		\bibfield  {author} {\bibinfo {author} {\bibfnamefont {G.}~\bibnamefont
				{Prando}}, \bibinfo {author} {\bibfnamefont {J.}~\bibnamefont {Perego}},
			\bibinfo {author} {\bibfnamefont {M.}~\bibnamefont {Negroni}}, \bibinfo
			{author} {\bibfnamefont {M.}~\bibnamefont {Ricc\`o}}, \bibinfo {author}
			{\bibfnamefont {S.}~\bibnamefont {Bracco}}, \bibinfo {author} {\bibfnamefont
				{A.}~\bibnamefont {Comotti}}, \bibinfo {author} {\bibfnamefont
				{P.}~\bibnamefont {Sozzani}},\ and\ \bibinfo {author} {\bibfnamefont
				{P.}~\bibnamefont {Carretta}},\ }\bibfield  {title} {\bibinfo {title}
			{Molecular rotors in a metal-organic framework: muons on a hyper-fast
				carousel},\ }\href@noop {} {\bibfield  {journal} {\bibinfo  {journal} {Nano
					Lett.}\ }\textbf {\bibinfo {volume} {20}},\ \bibinfo {pages} {7613} (\bibinfo
			{year} {2020}{\natexlab{b}})}\BibitemShut {NoStop}%
		\bibitem [{\citenamefont {Carretta}\ and\ \citenamefont
			{Prando}(2020)}]{Carretta2020RNC}%
		\BibitemOpen
		\bibfield  {author} {\bibinfo {author} {\bibfnamefont {P.}~\bibnamefont
				{Carretta}}\ and\ \bibinfo {author} {\bibfnamefont {G.}~\bibnamefont
				{Prando}},\ }\bibfield  {title} {\bibinfo {title} {Iron-based
				superconductors: tales from the nuclei},\ }\href@noop {} {\bibfield
			{journal} {\bibinfo  {journal} {Riv. Nuovo Cimento}\ }\textbf {\bibinfo
				{volume} {43}},\ \bibinfo {pages} {1} (\bibinfo {year} {2020})}\BibitemShut
		{NoStop}%
		\bibitem [{\citenamefont {Prando}\ \emph {et~al.}(2011)\citenamefont {Prando},
			\citenamefont {Carretta}, \citenamefont {De~Renzi}, \citenamefont {Sanna},
			\citenamefont {Palenzona}, \citenamefont {Putti},\ and\ \citenamefont
			{Tropeano}}]{Prando2011}%
		\BibitemOpen
		\bibfield  {author} {\bibinfo {author} {\bibfnamefont {G.}~\bibnamefont
				{Prando}}, \bibinfo {author} {\bibfnamefont {P.}~\bibnamefont {Carretta}},
			\bibinfo {author} {\bibfnamefont {R.}~\bibnamefont {De~Renzi}}, \bibinfo
			{author} {\bibfnamefont {S.}~\bibnamefont {Sanna}}, \bibinfo {author}
			{\bibfnamefont {A.}~\bibnamefont {Palenzona}}, \bibinfo {author}
			{\bibfnamefont {M.}~\bibnamefont {Putti}},\ and\ \bibinfo {author}
			{\bibfnamefont {M.}~\bibnamefont {Tropeano}},\ }\bibfield  {title} {\bibinfo
			{title} {Vortex dynamics and irreversibility line in optimally doped
				{SmFeAsO$_{0.8}$F$_{0.2}$} from ac susceptibility and magnetization
				measurements},\ }\href@noop {} {\bibfield  {journal} {\bibinfo  {journal}
				{Phys. Rev. B}\ }\textbf {\bibinfo {volume} {83}},\ \bibinfo {pages} {174514}
			(\bibinfo {year} {2011})}\BibitemShut {NoStop}%
		\bibitem [{\citenamefont {Prando}\ \emph {et~al.}(2012)\citenamefont {Prando},
			\citenamefont {Carretta}, \citenamefont {De~Renzi}, \citenamefont {Sanna},
			\citenamefont {Grafe}, \citenamefont {Wurmehl},\ and\ \citenamefont
			{B\"uchner}}]{Prando2012}%
		\BibitemOpen
		\bibfield  {author} {\bibinfo {author} {\bibfnamefont {G.}~\bibnamefont
				{Prando}}, \bibinfo {author} {\bibfnamefont {P.}~\bibnamefont {Carretta}},
			\bibinfo {author} {\bibfnamefont {R.}~\bibnamefont {De~Renzi}}, \bibinfo
			{author} {\bibfnamefont {S.}~\bibnamefont {Sanna}}, \bibinfo {author}
			{\bibfnamefont {H.-J.}\ \bibnamefont {Grafe}}, \bibinfo {author}
			{\bibfnamefont {S.}~\bibnamefont {Wurmehl}},\ and\ \bibinfo {author}
			{\bibfnamefont {B.}~\bibnamefont {B\"uchner}},\ }\bibfield  {title} {\bibinfo
			{title} {ac susceptibility investigation of vortex dynamics in nearly
				optimally doped {$R$FeAsO$_{1-x}$F$_{x}$} superconductors ({$R$ = La, Ce,
					Sm})},\ }\href@noop {} {\bibfield  {journal} {\bibinfo  {journal} {Phys. Rev.
					B}\ }\textbf {\bibinfo {volume} {85}},\ \bibinfo {pages} {144522} (\bibinfo
			{year} {2012})}\BibitemShut {NoStop}%
		\bibitem [{\citenamefont {Prando}\ \emph {et~al.}(2013)\citenamefont {Prando},
			\citenamefont {Giraud}, \citenamefont {Aswartham}, \citenamefont {Vakaliuk},
			\citenamefont {Abdel-Hafiez}, \citenamefont {Hess}, \citenamefont {Wurmehl},
			\citenamefont {Wolter},\ and\ \citenamefont {B\"uchner}}]{Prando2013}%
		\BibitemOpen
		\bibfield  {author} {\bibinfo {author} {\bibfnamefont {G.}~\bibnamefont
				{Prando}}, \bibinfo {author} {\bibfnamefont {R.}~\bibnamefont {Giraud}},
			\bibinfo {author} {\bibfnamefont {S.}~\bibnamefont {Aswartham}}, \bibinfo
			{author} {\bibfnamefont {O.}~\bibnamefont {Vakaliuk}}, \bibinfo {author}
			{\bibfnamefont {M.}~\bibnamefont {Abdel-Hafiez}}, \bibinfo {author}
			{\bibfnamefont {C.}~\bibnamefont {Hess}}, \bibinfo {author} {\bibfnamefont
				{S.}~\bibnamefont {Wurmehl}}, \bibinfo {author} {\bibfnamefont {A.~U.~B.}\
				\bibnamefont {Wolter}},\ and\ \bibinfo {author} {\bibfnamefont
				{B.}~\bibnamefont {B\"uchner}},\ }\bibfield  {title} {\bibinfo {title}
			{Evidence for a vortex-glass transition in superconducting
				{Ba(Fe$_{0.9}$Co$_{0.1}$)$_{2}$As$_{2}$}},\ }\href@noop {} {\bibfield
			{journal} {\bibinfo  {journal} {J. Phys. Condens. Matter}\ }\textbf {\bibinfo
				{volume} {25}},\ \bibinfo {pages} {505701} (\bibinfo {year}
			{2013})}\BibitemShut {NoStop}%
		\bibitem [{\citenamefont {Vlasenko}\ \emph {et~al.}(2020)\citenamefont
			{Vlasenko}, \citenamefont {Pervakov},\ and\ \citenamefont
			{Gavrilkin}}]{Vlasenko2020sust}%
		\BibitemOpen
		\bibfield  {author} {\bibinfo {author} {\bibfnamefont {V.}~\bibnamefont
				{Vlasenko}}, \bibinfo {author} {\bibfnamefont {K.}~\bibnamefont {Pervakov}},\
			and\ \bibinfo {author} {\bibfnamefont {S.}~\bibnamefont {Gavrilkin}},\
		}\bibfield  {title} {\bibinfo {title} {Vortex pinning and magnetic phase
				diagram of {EuRbFe$_4$As$_4$} iron-based superconductor},\ }\href@noop {}
		{\bibfield  {journal} {\bibinfo  {journal} {Supercond. Sci. Technol.}\
			}\textbf {\bibinfo {volume} {33}},\ \bibinfo {pages} {084009} (\bibinfo
			{year} {2020})}\BibitemShut {NoStop}%
		\bibitem [{\citenamefont {Wang}\ and\ \citenamefont
			{Petrovic}(2017)}]{Wang2017APL}%
		\BibitemOpen
		\bibfield  {author} {\bibinfo {author} {\bibfnamefont {A.}~\bibnamefont
				{Wang}}\ and\ \bibinfo {author} {\bibfnamefont {C.}~\bibnamefont
				{Petrovic}},\ }\bibfield  {title} {\bibinfo {title} {Vortex pinning and
				irreversibility fields in {FeS$_{1-x}$Se$_x$} ($x = 0, 0.06$)},\ }\href@noop
		{} {\bibfield  {journal} {\bibinfo  {journal} {Appl. Phys. Lett.}\ }\textbf
			{\bibinfo {volume} {110}},\ \bibinfo {pages} {232601} (\bibinfo {year}
			{2017})}\BibitemShut {NoStop}%
		\bibitem [{\citenamefont {Hebard}\ and\ \citenamefont
			{Fiory}(1980)}]{Hebard1980prl}%
		\BibitemOpen
		\bibfield  {author} {\bibinfo {author} {\bibfnamefont {A.~F.}\ \bibnamefont
				{Hebard}}\ and\ \bibinfo {author} {\bibfnamefont {A.~T.}\ \bibnamefont
				{Fiory}},\ }\bibfield  {title} {\bibinfo {title} {Evidence for the
				{Kosterlitz-Thouless} transition in thin superconducting aluminum films},\
		}\href@noop {} {\bibfield  {journal} {\bibinfo  {journal} {Phys. Rev. Lett.}\
			}\textbf {\bibinfo {volume} {44}},\ \bibinfo {pages} {291} (\bibinfo {year}
			{1980})}\BibitemShut {NoStop}%
		\bibitem [{\citenamefont {G\"om\"ory}(1997)}]{Gomory1997sust}%
		\BibitemOpen
		\bibfield  {author} {\bibinfo {author} {\bibfnamefont {F.}~\bibnamefont
				{G\"om\"ory}},\ }\bibfield  {title} {\bibinfo {title} {Characterization of
				high-temperature superconductors by {AC} susceptibility measurements},\
		}\href@noop {} {\bibfield  {journal} {\bibinfo  {journal} {Supercond. Sci.
					Technol.}\ }\textbf {\bibinfo {volume} {10}},\ \bibinfo {pages} {523}
			(\bibinfo {year} {1997})}\BibitemShut {NoStop}%
		\bibitem [{\citenamefont {Belkin}\ \emph {et~al.}(2008)\citenamefont {Belkin},
			\citenamefont {Novosad}, \citenamefont {Iavarone}, \citenamefont {Pearson},\
			and\ \citenamefont {Karapetrov}}]{Belkin2008prb}%
		\BibitemOpen
		\bibfield  {author} {\bibinfo {author} {\bibfnamefont {A.}~\bibnamefont
				{Belkin}}, \bibinfo {author} {\bibfnamefont {V.}~\bibnamefont {Novosad}},
			\bibinfo {author} {\bibfnamefont {M.}~\bibnamefont {Iavarone}}, \bibinfo
			{author} {\bibfnamefont {J.}~\bibnamefont {Pearson}},\ and\ \bibinfo {author}
			{\bibfnamefont {G.}~\bibnamefont {Karapetrov}},\ }\bibfield  {title}
		{\bibinfo {title} {Superconductor/ferromagnet bilayers: influence of magnetic
				domain structure on vortex dynamics},\ }\href@noop {} {\bibfield  {journal}
			{\bibinfo  {journal} {Phys. Rev. B}\ }\textbf {\bibinfo {volume} {77}},\
			\bibinfo {pages} {180506} (\bibinfo {year} {2008})}\BibitemShut {NoStop}%
		\bibitem [{\citenamefont {Lange}\ \emph {et~al.}(2005)\citenamefont {Lange},
			\citenamefont {Van~Bael}, \citenamefont {Silhanek},\ and\ \citenamefont
			{Moshchalkov}}]{Lange2005prb}%
		\BibitemOpen
		\bibfield  {author} {\bibinfo {author} {\bibfnamefont {M.}~\bibnamefont
				{Lange}}, \bibinfo {author} {\bibfnamefont {M.~J.}\ \bibnamefont {Van~Bael}},
			\bibinfo {author} {\bibfnamefont {A.~V.}\ \bibnamefont {Silhanek}},\ and\
			\bibinfo {author} {\bibfnamefont {V.~V.}\ \bibnamefont {Moshchalkov}},\
		}\bibfield  {title} {\bibinfo {title} {Vortex-antivortex dynamics and
				field-polarity-dependent flux creep in hybrid superconductor/ferromagnet
				nanostructures},\ }\href@noop {} {\bibfield  {journal} {\bibinfo  {journal}
				{Phys. Rev. B}\ }\textbf {\bibinfo {volume} {72}},\ \bibinfo {pages} {052507}
			(\bibinfo {year} {2005})}\BibitemShut {NoStop}%
	\end{thebibliography}
\end{document}